\documentclass[preprint,showpacs,aps,a4paper]{revtex4}

\usepackage{graphicx}
\usepackage{dcolumn}
\usepackage{bm}
\usepackage[english,spanish,activeacute]{babel}
\def\journal#1#2#3#4{{\sl #1,\/} {\bf #2}, #3 (#4)}

\begin{document}
\selectlanguage{english}

\title{The influence of the  geomagnetic field and of the 
 uncertainties in the primary spectrum on the development
 of the muon flux in the atmosphere. 
}
\author{P.~Hansen, T.~K.~Gaisser, T. ~Stanev}
\affiliation{Bartol Research Institute\\
             University of Delaware\\
             Newark, Delaware 19716,\\
             USA.}
\author{S.~J.~Sciutto}
\affiliation{Departamento de F\'isica and IFLP/CONICET\\
             Universidad Nacional de La Plata\\
             C. C. 67 - 1900 La Plata\\
             Argentina.}

\date{\today}

\begin{abstract}

 In this paper we study the sensitivity of the flux of
 atmospheric muons to uncertainties in the
 primary cosmic ray spectrum and to the treatment of the
 geomagnetic field in a calculation.
 We use the air shower simulation program AIRES to make
 the calculation for two different primary
 spectra and under several approximations to the propagation 
 of charged particles in the geomagnetic field. 
 The results illustrate the importance of
 accurate modelling of the geomagnetic field effects.
 We propose a high and a low fit of the proton and helium fluxes,
 and calculate the muon fluxes with these different inputs.
 Comparison with measurements
 of the muon flux by the CAPRICE experiment shows a slight preference
 for the higher primary cosmic ray flux parametrization.

\end{abstract}

\maketitle

\section{Introduction}

 The study of the muon fluxes in the atmosphere is currently 
 of great interest because of the correlation between the muon 
 and neutrino fluxes.
 Atmospheric neutrinos are produced from the decay channels of 
 pions and kaons and the subsequent muon decay.
 The production of electron and muon neutrinos is dominated by
 the processes $\pi ^{+}\rightarrow \mu ^{+}+\nu _{\mu }$
 followed by $\mu ^{+}\rightarrow e^{+}+\overline{\nu}_{\mu }+\nu _{e}$
 (and their charge conjugates), with a similar chain for charged kaons.
 When  all particles decay, there will be two muon neutrinos for
 each electron neutrino resulting in an expected ratio
 $ \nu _{\mu}/\nu _{e}$
 of the flux of $\nu _{\mu }+\overline{\nu }_{\mu }$ 
 to the flux of $\nu _{e}+\overline{\nu} _{e}$ of about 2.
 The experimental measurements \cite{k1,k2,imb,sk1,frejus,nusex,ultima} 
 indicate, however, that the ratio of muon to electron neutrinos
 in the atmosphere is significantly smaller than two.

 This disagreement with theoretical predictions has been interpreted
 in terms of neutrino oscillations \cite{oscila}.
 In order to calculate precisely the neutrino oscillation parameters
 one has to know the neutrino flux at production.
 Because of the close connection beteween the neutrinos and the muons,
a standard test of codes used for calculating the neutrino flux is to
calculate the muon flux with the same procedure and compare it
to measurements of muons.  Our approach here is related but somewhat
different.  We calculate only the muon flux, and we use the comparison
with measurements to probe two aspects of the input that are common to
both the neutrino and the muon fluxes.

 The calculation of both muons and neutrinos starts
 with the primary spectrum outside the atmosphere.
Therefore uncertainties in the measurements of the primary spectrum of
 protons, helium and heavier nuclei affect both fluxes in a similar way.
Because the uncorrelated fluxes depend essentially on energy per nucleon,
heavy nuclei have relatively little effect, and the largest uncertainty
in normalization comes from protons and helium.
Experimental measurements of primary proton and helium spectra
in some cases show significant differences from each other.
We find that it is possible to divide the experimental data over a fairly
large energy range into two groups:
group 1 corresponds to the data wich give a lower flux
and group 2 to a higher flux.  By making a ``high" fit and
a ``low" fit over an extended energy range, we can define
a reasonable range where the primary spectum should lie.
Comparison with measured muon fluxes may provide
an extra constraint on the normalization of the primary flux.

Treatment of the geomagnetic field affects both the neutrino and
the muon fluxes.  One effect is a consequence of the field acting on
the primary cosmic rays, which determines allowed and forbidden
trajectories.  Primaries on allowed trajectories reach the
atmosphere to interact and produce secondary muons and neutrinos while
those on forbidden trajectories do not reach the atmosphere and therefore
do not contribute to secondary fluxes.  The other significant effect
is the bending of charged muons after production in the atmosphere.
Which trajectories are allowed and which forbidden depends both on
magnetic rigidity (defined as total momentum divided by charge of the
nucleus) and on the direction of the particle.  At high geomagnetic 
latitudes all primaries with energies above pion production threshold
are allowed.  At low latitudes, particles need to have a minimum
rigidity to reach the atmosphere, and this minimum value is higher
for positive particles from the East than from the West.  We 
investigate the sensitivity of the muon fluxes
to both aspects of the geomagnetic field as a function energy and
atmospheric depth. 

 The paper is organized as follows: 
 Section \ref{S:Thesimulations} is divided in three subsections.
 In \ref{S:Airshowers} we report on the method used to
 calculate the atmospheric muon fluxes; in \ref{S:seast}
 we describe the treatment of the geomagnetic 
 field; and in \ref{S:input} we propose the 
 high and low fits of the primary proton and helium spectrum data.
 The results are reported in \ref{S:results} and finally 
 we summarize our conclusions in section \ref{S:conclusion}.

\section{The simulations}
\label{S:Thesimulations}

\subsection{Air shower simulations}
\label{S:Airshowers}
 In this work we have used the Air Shower Simulation Program (AIRES)
 \cite{AIRES,AIRESManual}, which provides
 full space-time particle propagation and accounts for 
 the atmospheric density profile \cite{atmosfera} and the Earth's
 curvature.
 The particles taken into account by AIRES in the simulation
 are: gammas, electrons, positrons, muons, pion, kaons,
 lambda baryons, nucleons, antinucleons and nuclei
 up to Z=36. The hadronic processes are simulated
 using different models, according with their energy.
 High energy collisions are processed invoking an external package,
 SIBYLL or QGSjet (SIBYLL
 2.1 \cite{SIBYLL} or QGSjet01 \cite{QGSJET}), while low energy
 ones are processed using an extension
 of the Hillas splitting algorithm (EHSA)\cite{AIRES,AIRESCCICRC,AUGERJPG}.
 In this work we use SIBYLL. The threshold energy separating the
 low and high energy regimes is 200 GeV.
 The effect of the geomagnetic field (GF) 
inside the atmosphere is taken into account
 in AIRES.
 The GF calculations are controlled from the input instruction
 by specifying a date and the geographic coordinates of a site.
 The program uses the IGRF model \cite{GF}
to evaluate the magnetic field intensity and orientation.
 It is assumed that the shower develops under the influence of
 a constant and homogeneous field which is evaluated before starting
 the simulations.

 The input used in the simulation and the procedure to calculate the 
 flux is the same as that used in Ref \cite{sergioyyo}.
In Ref~\cite{sergioyyo} the  CAPRICE experimental geomagnetic
 transmission function was used to estimate the cutoff prior to
 the  calculation of the muon flux.  Here we use a more precise
backtracking method, as described in the next subsection.

\subsection{Geomagnetic field effects}
\label{S:seast}

 The magnetic field affects the low energy muon flux
 both through the geomagnetic cutoffs on the primary
 cosmic rays, including the East-West effect, and by
 the bending of trajectories of secondary charged particles
 inside the atmosphere.

 The East-West effect is the suppression of cosmic ray nuclei incident
 on the atmosphere from the East compared to those from the West.
 This suppression is due to the combination of the following two 
 facts \cite{johnson} \cite{alvarez}:
\begin{description}
\item[-] Positively charged particles at the same zenith angle have
 a higher cutoff from the East direction  than from the West (and
 vice-versa for negatively
 charged particles) since some of their trajectories intersect the Earth.
\item[-] Cosmic rays are positively charged nuclei, so they
 will bend in one sense in the geomagnetic field.
\end{description}

 To calculate these geomagnetic effects we use backtracking
 technique in a detailed, time-dependent geomagnetic field model
 (IGRF, International Geomagnetic Field model).
 This technique consists of the integration of 
 the equation of motion of a particle with the opposite charge
 starting at a position near the top of the atmosphere.
 We inject antiprotons outwards from an altitude of 100 km 
 in various directions and see if the backtracked antiproton reaches
 a distance of 30 R$_\oplus$ from the Earth within a total pathlength
 of 300 R$_\oplus$. Any direction in which an antiproton of a
 given momentum can reach this distance is an allowed direction
 from which a proton of the opposite momentum can arrive.
 The backtracked antiprotons that do not reach that distance
 are either trapped in the geomagnetic field or their trajectories
 intersect the surface of the Earth. In these last two cases the
 trajectory is considered forbidden.
 The results are expressed in terms of a transmission
 function, also called penetration probability, which is  
 zero at low rigidity and increases to one at higher rigidity.
 The energy dependence of the transmission function depends on the
 geomagnetic latitude and the angle between the cosmic ray direction 
 and the geomagnetic field lines.
 The geomagnetic cutoff is treated in the simulation by the
 application of this function to the input primary cosmic ray flux,

 To explore the East-West effect we have compared two transmission
 functions, one averaged over primaries inside a cone of half angle 
 30$^\circ$
 from the East and the other over a similar cone from the West.

The CAPRICE Geomagnetic transmission function \cite{emiliano} used
in Ref.~\cite{sergioyyo} was obtained by comparing the
 shape of the spectra of alpha particles measured by
 the balloon borne experiment CAPRICE94~\cite{boe99b} with the shape
 of CAPRICE98~\cite{amb99}.
 These two balloon experiments flew in different locations:
 the first in Lynn Lake, Manitoba, Canada, where the geomagnetic
 cutoff is negligible,  and the other in Ft. Summer New Mexico (USA)
 where the vertical cutoff is 4.3 GV.
 This transmission function is only a function of rigidity and therefore
 does not produce the East-West effect.  We compare the two methods in 
 Section~\ref{S:results} below.

\subsection{The low and high fits to the primary Spectrum data}
\label{S:input}
 New measurements of the primary spectra of protons
 and helium have improved our knowledge of the primary 
 spectrum up to 100 GeV compared to what was previously known.
 There are nevertheless still significant discrepancies
 between different experiments. For this reason we have performed
 two different fits: one with the experiments that give the lowest 
 (we will call them Group 1) and another
 that give the highest fluxes (Group 2).
 All these measurements were fitted to the following function:

\begin{equation}
\phi (E_{k})=k \times (E_{k}+b\phantom{0}exp\left(-c \sqrt E_{k} \right))^{-\alpha}
\label{equ:stanevfunction}
\end{equation}

where $E_{k}$ is the kinetic energy per nucleon~\cite{hamburgcrf}, and
$\alpha$ and $k$ are free parameters.

 Group 1 consists of the data of CAPRICE98 \cite{emiliano}, 
 Atic \cite{ATICi}  \cite{ATIC}  below 100 GeV and
 RunJob~\cite{RunJOB} \cite{RunJOBi} at high energy.
 Group 2 consists of the data of AMS \cite{AMSp}~\cite{AMSh},
 BESS \cite{Bess} at low energy and JACEE \cite{Jacee} at high energy
 (There was not a considerable
 difference between AMS-JACEE and BESS-JACEE fit for the protons
 fluxes).

\begin{figure}[ht]
{\centering \begin{tabular}{cc}
\includegraphics[angle=0, width=8.0cm]{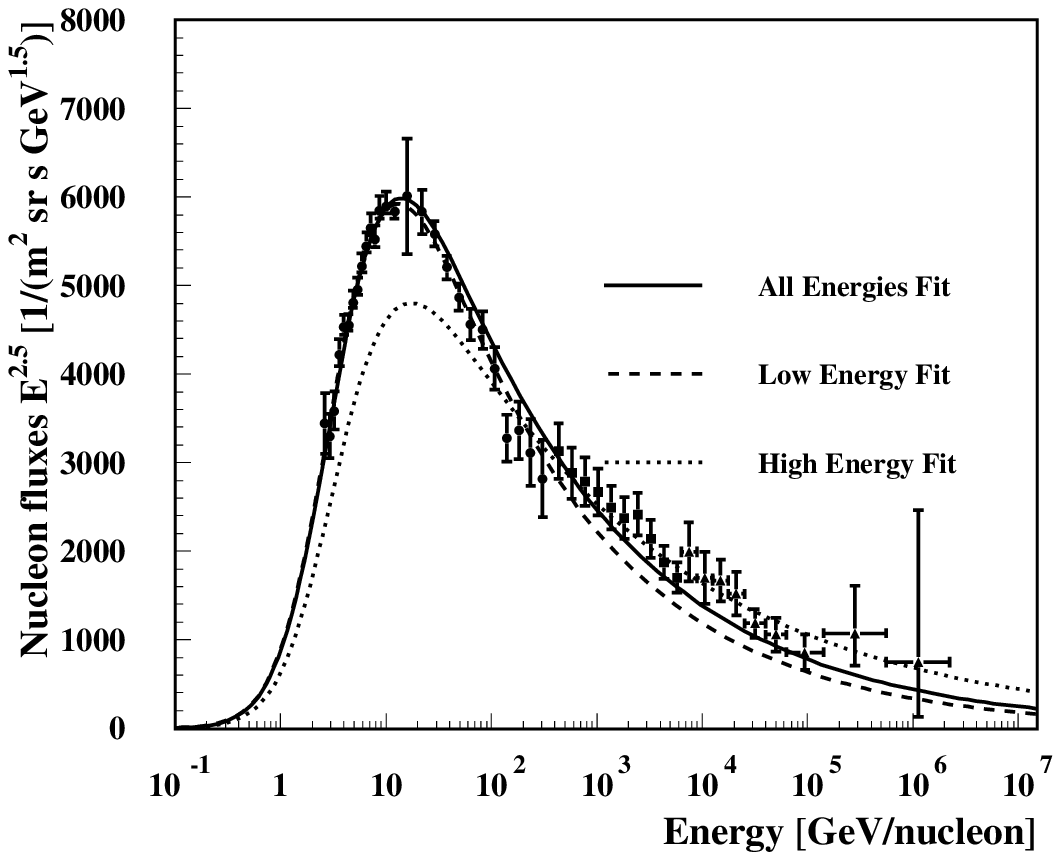}
&\includegraphics[angle=0, width=8.0cm]{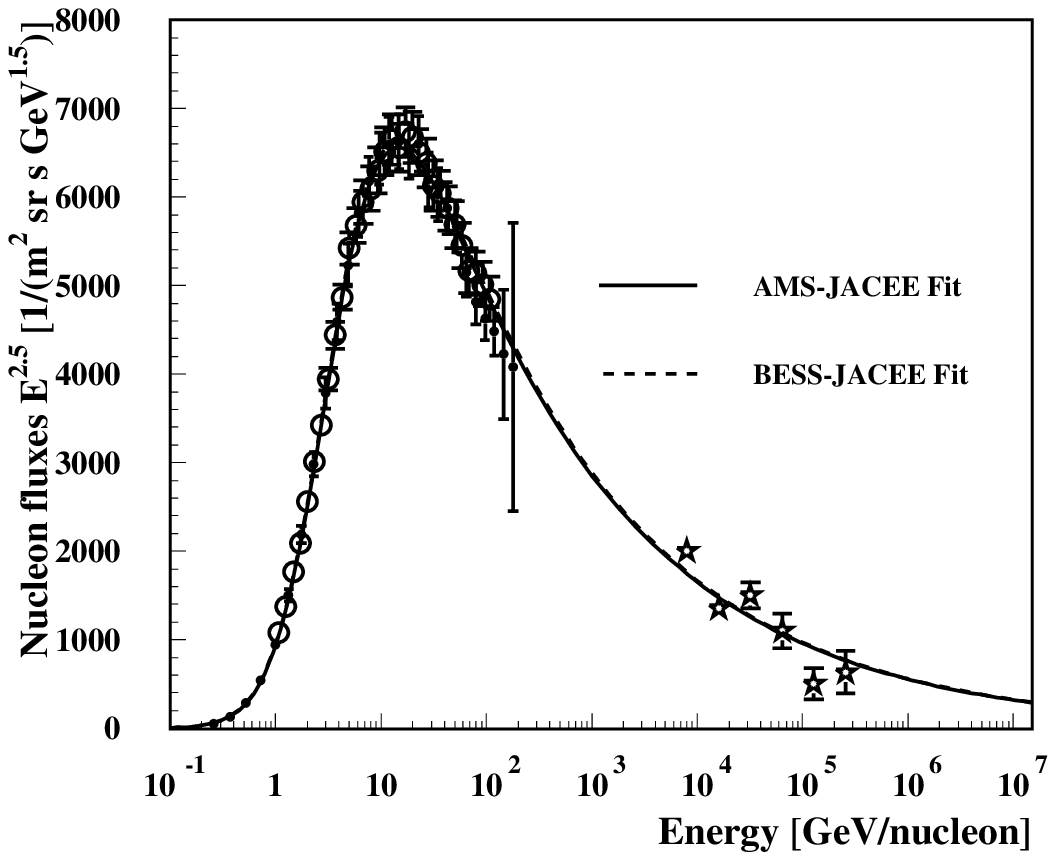}
\\
\end{tabular}\par}
\caption{\textbf{Left:} Proton fluxes from CAPRICE98 (full circle) Atic
(black square) and RunJob (black triangle)
(group 1) as function
of the energy per nucleon.
The lines correspond to the different fits of the data.
\textbf{Right:} Proton fluxes data of
AMS (full circle) - JACEE (star)   and BESS (open circle) - JACEE (star)
(group 2) as function of the energy per nucleon.
The lines correspond to the different fits of the data.
}
\label{fig:fig1}
\end{figure}

  In figure~\ref{fig:fig1} we show the
  combination of the experimental data on the proton fluxes
  for the experiments of group 1 (left) and for group 2 (right).
  Three different fits are shown in the left-hand panel.
  The solid line corresponds to the fit of all the data.
  We have also made separate fits for the low energy data
  (the dashed line) and for the high energy data (the dotted line).
  The left hand side of the figure~\ref{fig:fig1} shows the proton 
  data from AMS with full circles, BESS with open circles and JACEE
  with stars.
  We have done two fits for the following two groups AMS-JACEE (solid line)
  and BESS-JACEE (dashed line).

\begin{figure}[ht]
{\centering \begin{tabular}{cc}
\includegraphics[angle=0, width=8.0cm]{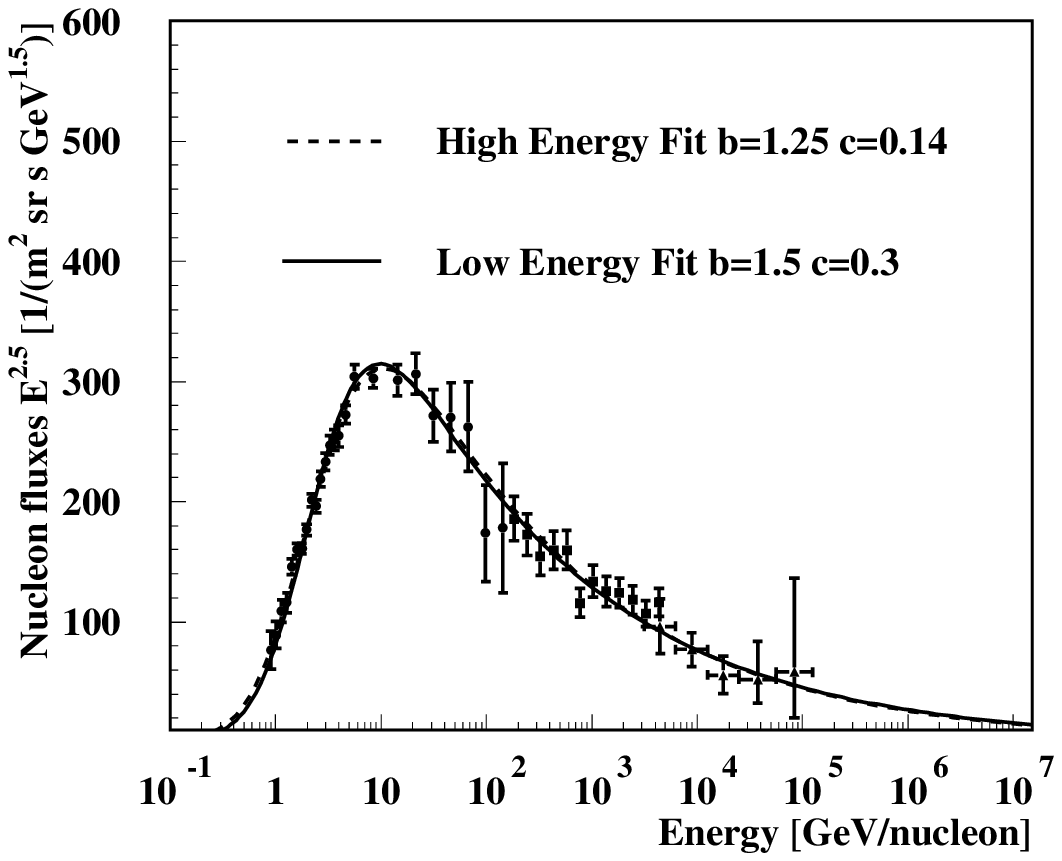}
&\includegraphics[angle=0, width=8.0cm]{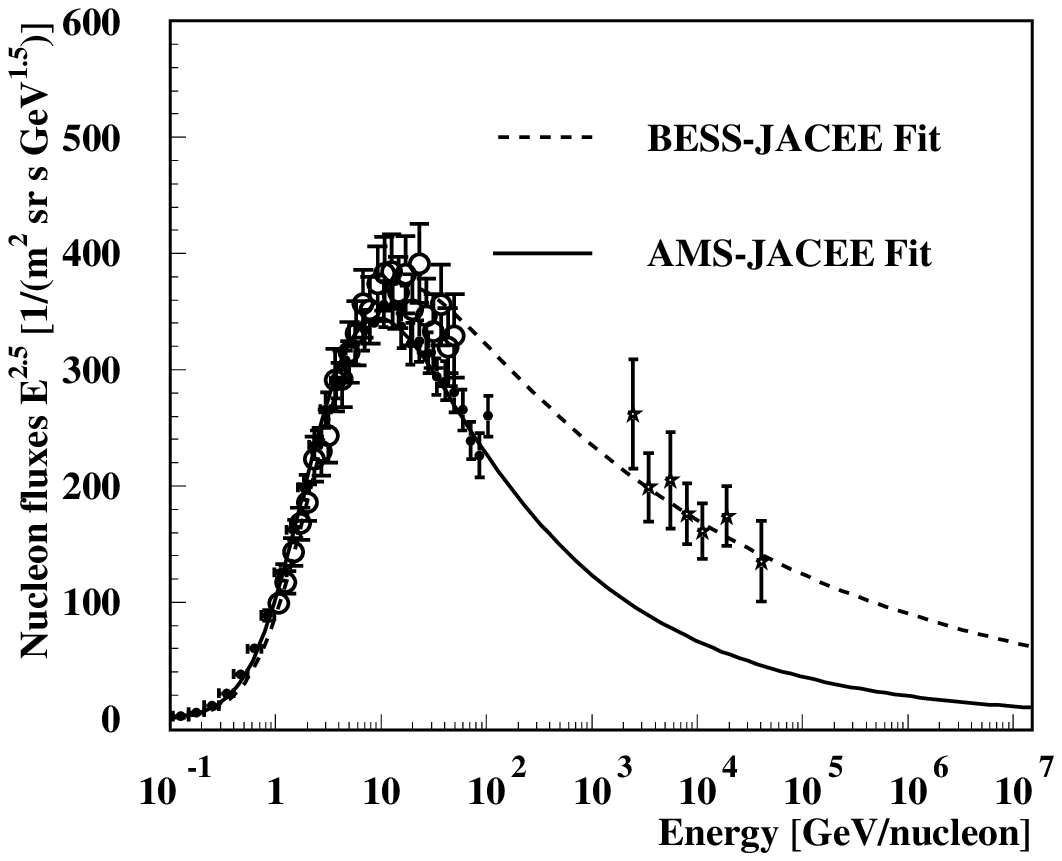}
\\
\end{tabular}\par}
\caption{\textbf{Left:} Helium fluxes from CAPRICE98 (full circle), Atic
(black square)  and RunJob (black triangle)
(group 1) as function
of the energy per nucleon.
The lines correspond to the different fits of the data.
\textbf{Right:} Helium fluxes data of
AMS (full circle) - JACEE  (star) and BESS (open circle)  - JACEE (star) 
(group 2) as function of the energy per nucleon.
The lines correspond to the different fits of the data.
}
\label{fig:fig2}
\end{figure}

\begin{figure*}[ht]
{\begin{tabular}{cc}
\resizebox*{0.5\textwidth}{!}{\includegraphics{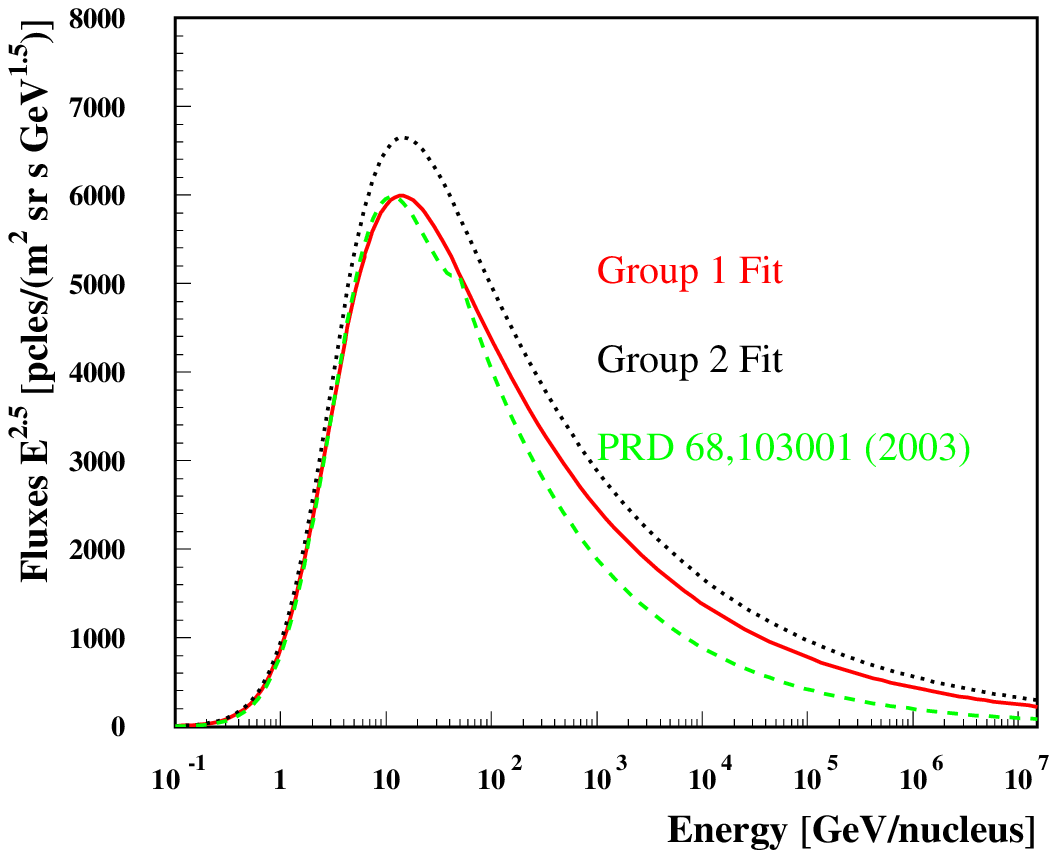}}
&
\resizebox*{0.5\textwidth}{!}{\includegraphics{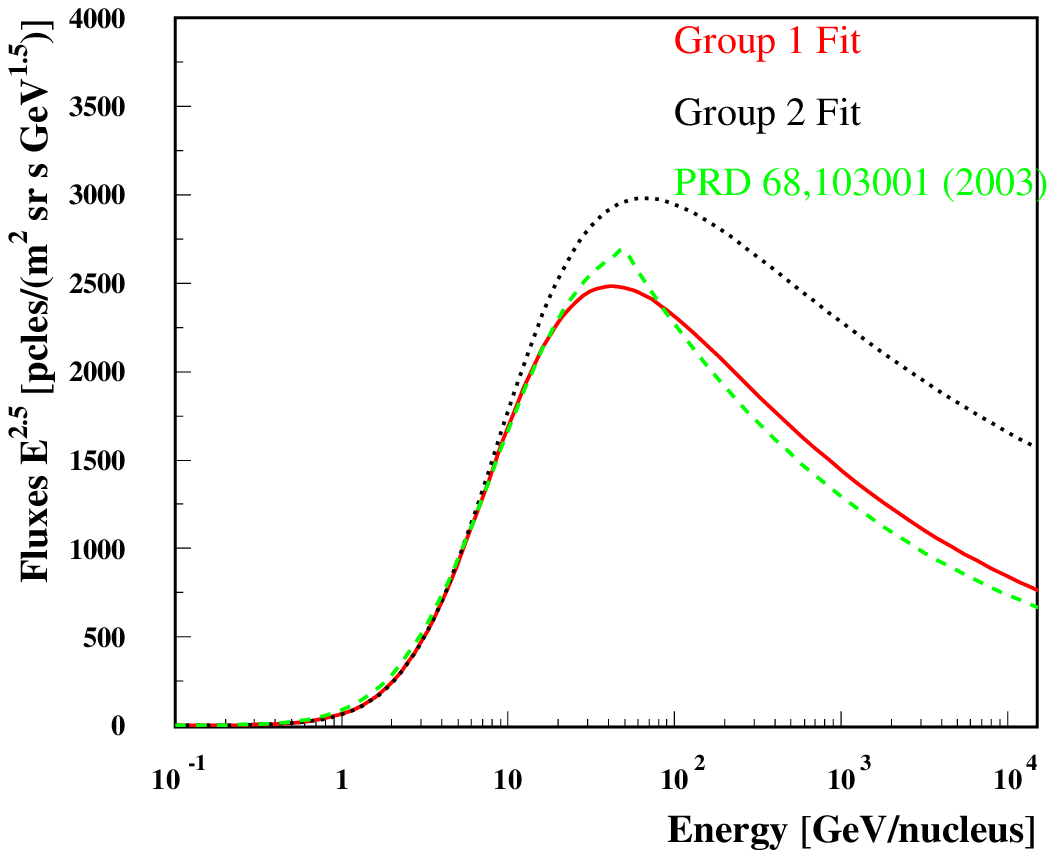}}
\\
\end{tabular}\par}
\caption{
Fluxes of proton (left) and helium (right) at the top
of the atmosphere, plotted as function of the kinetic energy
of the primaries.
{\em (1)\/} Solid line: Group 1 fit.
{\em (2)\/} Dotted line: Group 2 fit.
{\em (3)\/} Dashed line: Input used in Ref.~\cite{sergioyyo}.
}
\label{fig:fig3}
\end{figure*}

  Figure~\ref{fig:fig2} shows the helium experimental data
  for group 1 (left) and group 2 (right).
  For group 1 we have made two fits with different values of b and c.
  For group 2 there is a considerable difference between
  the data of AMS and JACEE. This is not the case in a combination of
  BESS and JACEE data. 
  For this reason we performed two different fits but
  implement in the calculation the BESS-JACEE data fit.
  The parameters for the fluxes of group 1 and group 2
  are shown in table~\ref{table:exp}.
  To have a clear picture 
  in figure~\ref{fig:fig3} we show three different fits  
  of the  absolute fluxes of proton (left) 
  and helium (right) that we will use in this calculation
  (Group 1, Group 2 and that of Ref.~\cite{sergioyyo}).

\begin{table}[htbp]
\caption{Parameters for the Hydrogen and Helium components in the fit
 of Equation \protect\ref{equ:stanevfunction} for the two data groups.
}
\begin{tabular}{|l|l|l|l|l|l|}
\hline
Group &\bf Component  & \bf \phantom{00000}$\alpha$   & \bf \phantom{0000} $k$  &\bf \phantom{0}$b$ &\bf \phantom{0} $c$\\
\hline
\phantom{00}1&Hydrogen & 2.751 $\pm$  0.004   &14000 $\pm$ 130 & 2.15 &0.21 \\
\phantom{00}1&Helium   & 2.734 $\pm$ 0.005  & 657 $\pm$  8  &1.25  &0.14  \\
\hline
\phantom{00}2&Hydrogen & 2.738 $\pm$  0.004   &15000 $\pm$ 160 & 2.15 &0.21 \\
\phantom{00}2&Helium   & 2.639 $\pm$ 0.008  & 615 $\pm$  16  &1.25  &0.14  \\
\hline
\end{tabular}
\label{table:exp}
\end{table}
 
 The fluxes of H and He were complemented by fluxes of heavy nuclei in
 eight groups that were fitted to the available data as in
 Ref.~\cite{sergioyyo}. The extensions of the spectra of heavy nuclei
 are not extremely important for the calculations of the atmospheric muon
 fluxes because  H and He nuclei provide 85 to 90\% of the all nucleon
 fluxes. The potential error from inexact fitting of the spectra of
 heavy nuclei would not exceed 3\% of the all nucleon flux.
 It is the normalization of the proton fluxes that dominate the difference
 in muon fluxes between group 1 and group 2.
 
\section{Results}
\label{S:results}
 
\subsection{Geomagnetic field effects.}

\begin{figure*}[ht]
{\centering \begin{tabular}{cc}
\includegraphics[width=8.0cm]{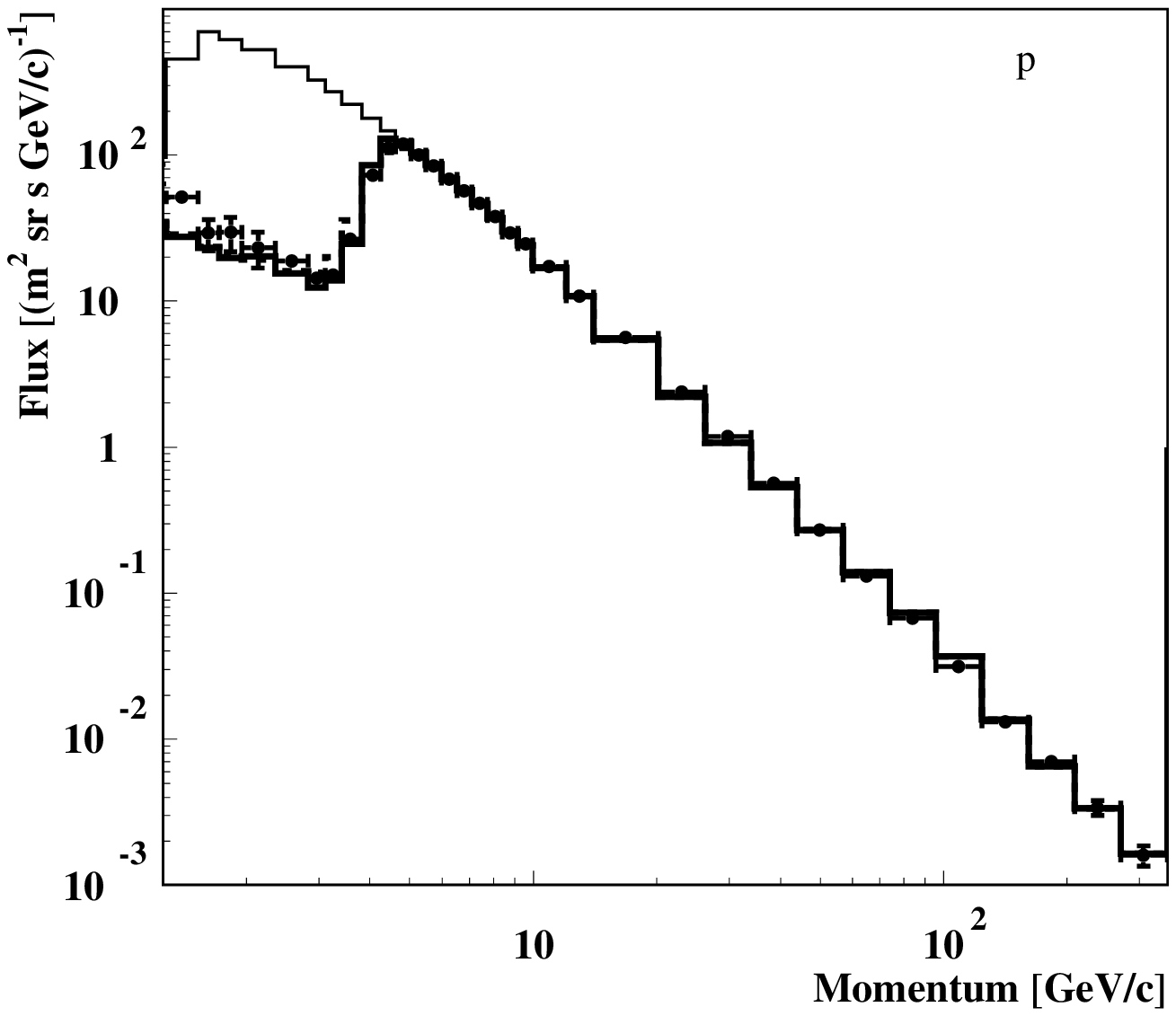}&
\includegraphics[width=8.0cm]{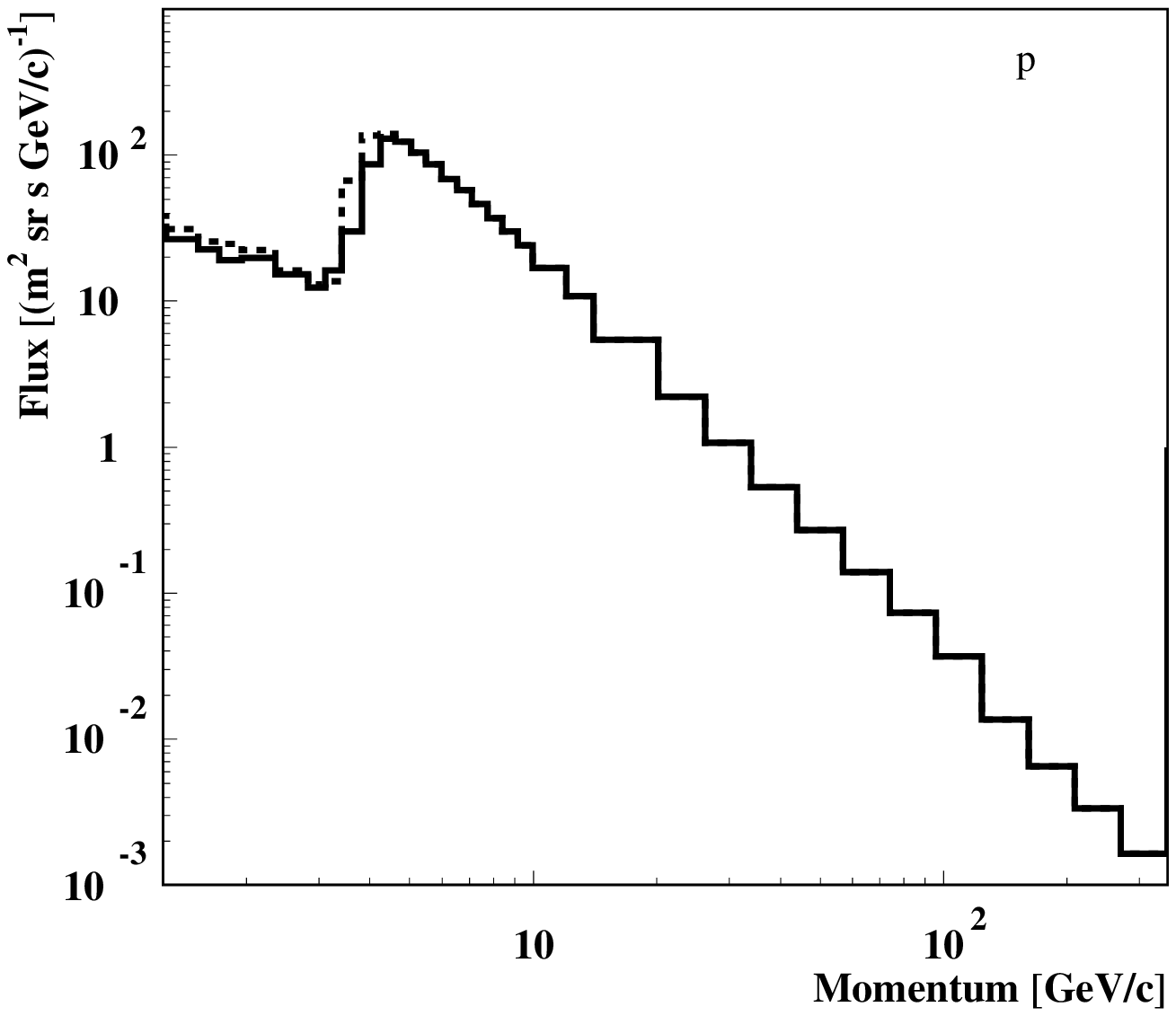}
\\
\end{tabular}\par}
\caption{\textbf{Left:}
The solid points ($\bullet$) represent the experimental proton flux
measured by CAPRICE98 at Fort Summer for a residual
atmospheric depth of
5.5 g/cm$^{2}$ (PhD thesis E. Mocchiutti, Royal Institute of Technology).
\newline
The lines are the simulated proton fluxes applying to the input spectrum:
\newline
{\em (1)\/} a theoretical geomagnetic transmission function (dotted line).
\newline
{\em (2)\/} the CAPRICE geomagnetic transmission function
(wide solid line).
\newline
{\em (3)\/} no geomagnetic transmission function (narrow solid line).
\newline
\textbf{Right:}
The lines are the simulated proton fluxes applying to the input spectrum:
\newline
{\em (1)\/} Dotted line: 
a theoretical geomagnetic transmission function that
also only select primary cosmic rays that came from the West.
\newline
{\em (2)\/} Wide solid line: 
a theoretical geomagnetic transmission function that
also only select primary cosmic rays that came from the East.
}
\label{fig:fig4}
\end{figure*}

 To check the validity of the transmission functions
 we have calculated the flux of protons at an altitude of
 5.5 g/cm$^{2}$ where we have experimental results from the CAPRICE98
 experiment. In figure~\ref{fig:fig4} the proton flux at
 5.5 g/cm$^{2}$ is plotted on the left as a
 function of momentum. The full points correspond to the
 data of the proton flux at 5.5 g/cm$^{2}$ from CAPRICE98
 experiment~\cite{emilianot} and the lines are the proton fluxes obtained
 by the AIRES simulation. The narrow solid line is
 without applying any transmission function, the wide solid line is
 using the CAPRICE transmission function and the dotted is
 applying the theoretical transmission function.
 We are able to reproduce quite well the proton flux at
 5.5 g/cm$^{2}$~\cite{emilianot} using both  the CAPRICE and
 the theoretical transmission functions.

 To illustrate the East-West effect we apply to our input
 flux the theoretical geomagnetic transmission function but
 selecting primaries particles that only
 came from the East or from the West.
 In figure~\ref{fig:fig4} on the right the dotted line
 corresponds to the flux of protons calculated using the
 theoretical geomagnetic transmission function from the West.
 The wide solid line corresponds to protons from the East.
 There is a definite excess of protons that come from the West
 at energies lower than 10 GeV.
 
\begin{figure*}[ht]
{\centering \begin{tabular}{c}
\includegraphics[angle=0, width=10.0cm]{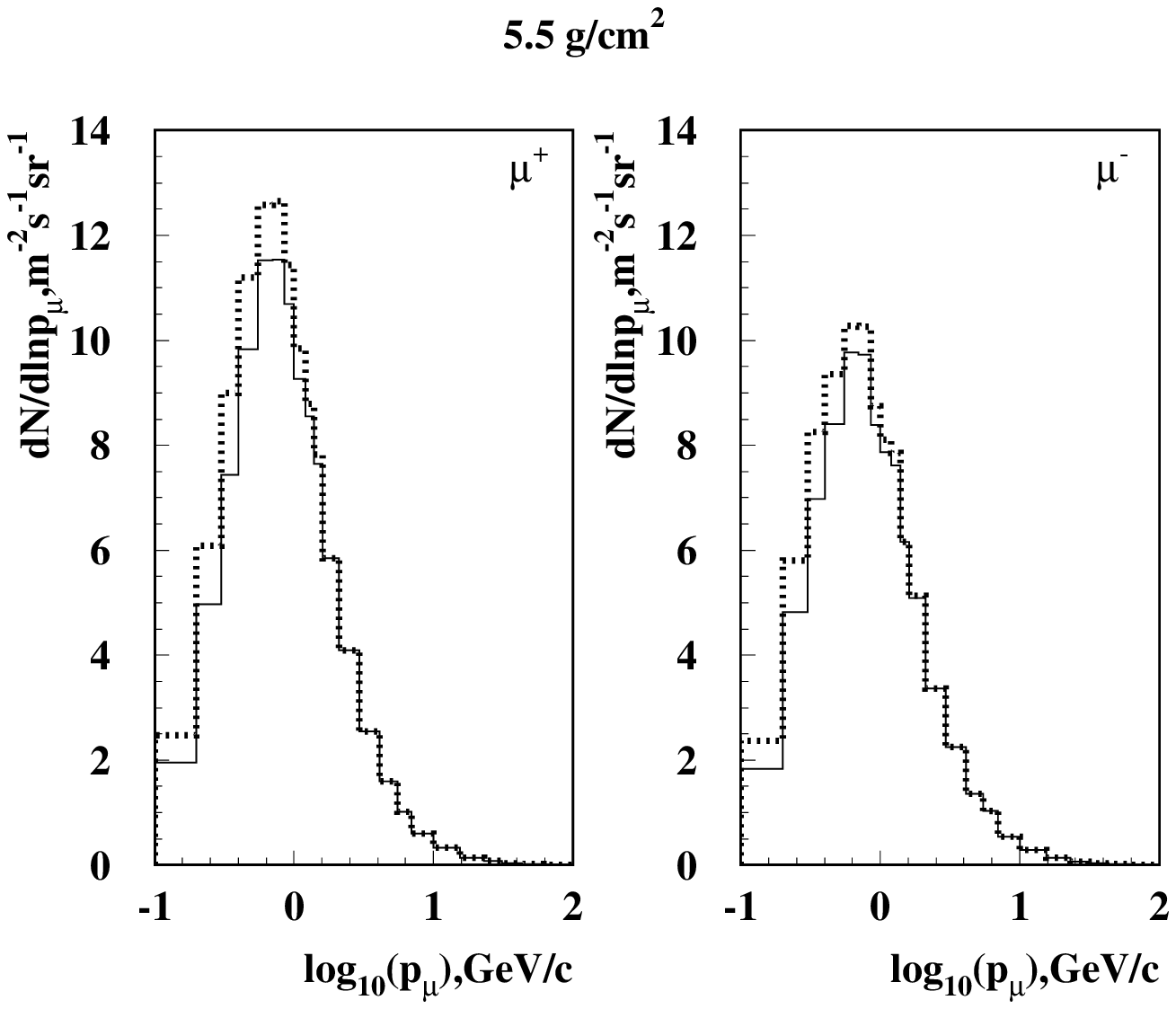}\\
\includegraphics[angle=0, width=10.0cm]{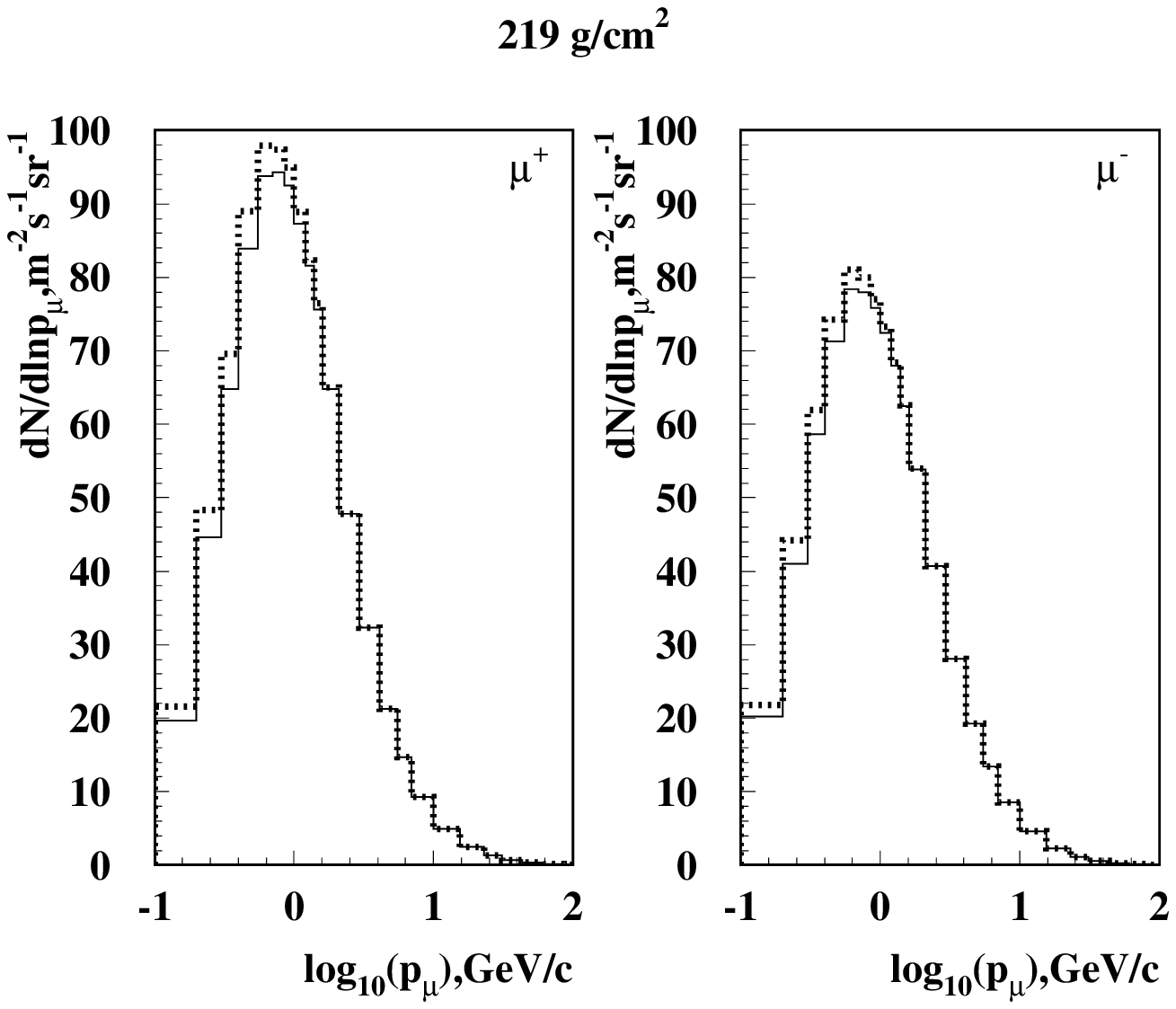}
\\
\end{tabular}\par}
\caption{Muon fluxes at Fort Summer for several
residual atmospheric depths applying the:
{\em (1)\/} Dotted line: Theoretical geomagnetic transmission function that
also only select primary cosmic rays that came from the West.
{\em (2)\/} Solid line: Theoretical geomagnetic transmission function that
also only select primary cosmic rays that came from the East.
}
\label{fig:fig5}
\end{figure*}

\begin{figure*}[ht]
{\centering \begin{tabular}{c}
\includegraphics[angle=0, width=10.0cm]{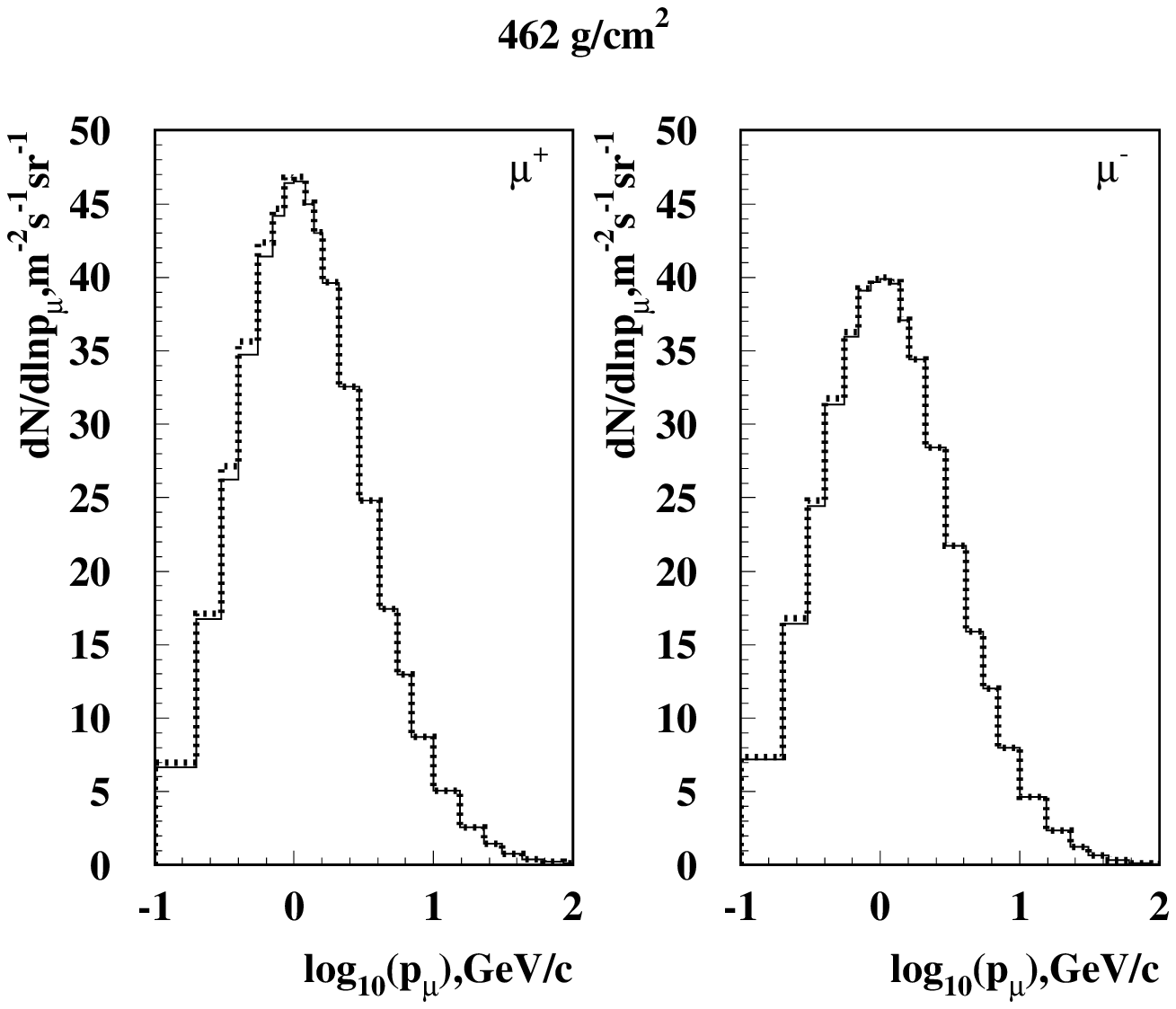}\\
\includegraphics[angle=0, width=10.0cm]{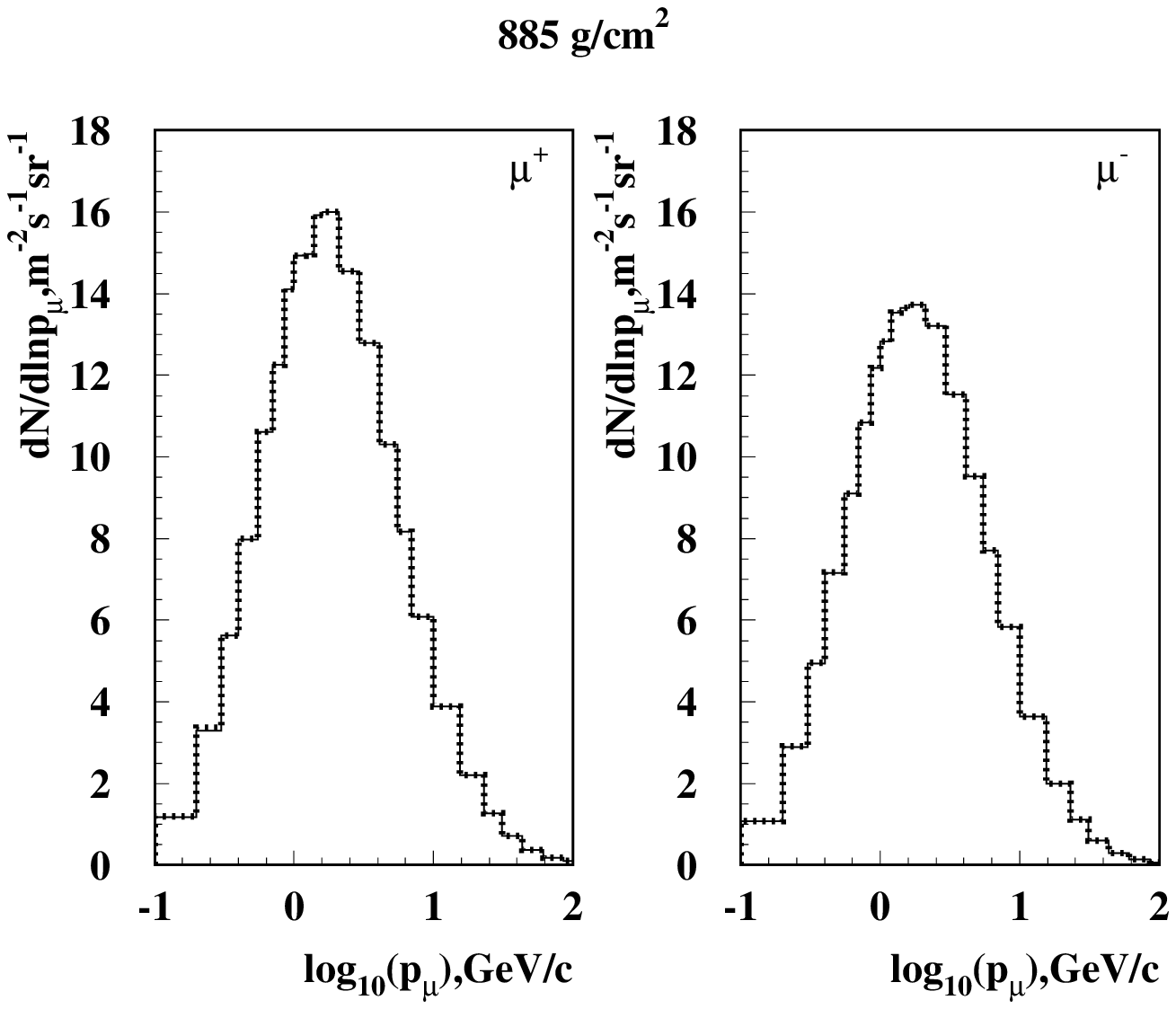}
\\
\end{tabular}\par}
\caption{Muon fluxes at Fort Summer for several
residual atmospheric depths applying the:
{\em (1)\/}  Dotted line: Theoretical geomagnetic transmission function that
also only select primary cosmic rays that came from the West.
{\em (2)\/} Solid line: Theoretical geomagnetic transmission function that
also only select primary cosmic rays that came from the East.
}
\label{fig:fig6}
\end{figure*}

 The effect of this excess on the muon fluxes  is illustrated
 in figures~\ref{fig:fig5}  and \ref{fig:fig6}.
 In these two figures we plot the muon flux coming from the West
 (dotted line) and coming from the East (solid line) as a function
 of the momentum.
 In the two figures it is possible to see that the excess
 of muons coming from the West decreases with increasing
 atmospheric depth and is negligible at the ground. To quantify the 
 results we calculate the relative differences between the East
 West fluxes for positives and negative muons. 
 It was found that the relative differences
 between the two directions for muons of energy around 0.2 GeV 
 is of order of 25 \% at  5.5 g/cm$^{2}$, 10 \% at 
 219 g/cm$^{2}$, 5 \% at 462  g/cm$^{2}$ and null at ground.

\begin{figure*}[ht]
\begin{center}
\includegraphics[width=12.0cm]{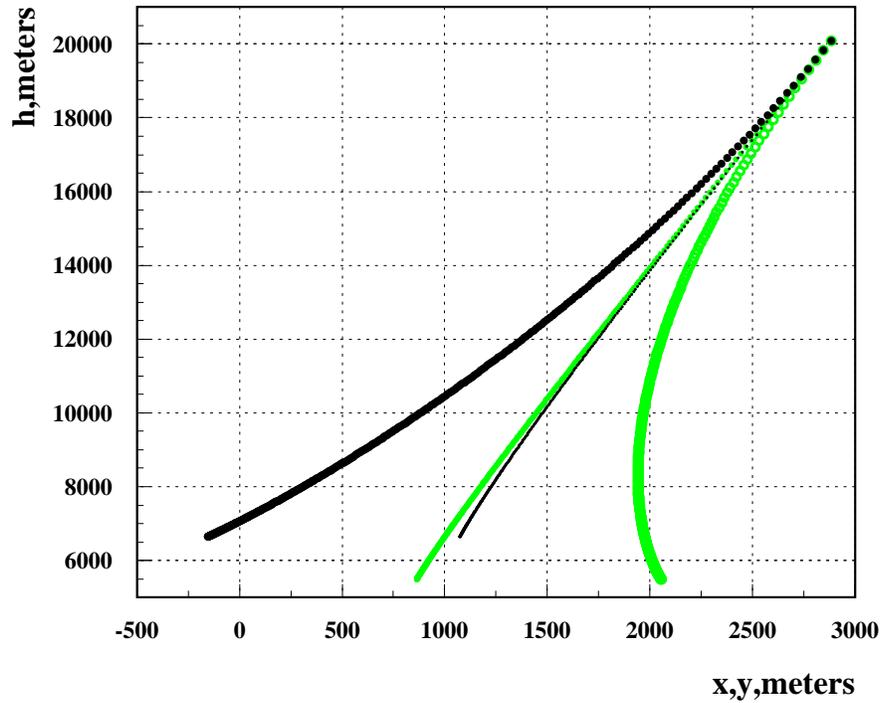}
\caption{A positive and a negative 1 GeV/c muon
injected at 20 Km with $cos(\theta)$=0.98 and azimuth=45$^{0}$
bending in the Geomagnetic field at For Sumner in x and y direction
 and followed to decay. 
{\em (1)\/} Narrow light line: Negative muon deflected in $x$ direction.
{\em (2)\/} Wide light line: Negative muon deflected in $y$ direction.
{\em (3)\/} Narrow solid line: Positive muon deflected in $x$ direction.
{\em (4)\/} Wide solid line: Positive muon deflected in $y$ direction.
The weight of the particle curves depend on the azimuth angle
and is stronger when the particle are perpendicular to $B$.}
\label{fig:fig7}
\end{center}
\end{figure*}

 The muon flux also changes because of the bending in the
 local magnetic field inside the atmosphere. 
 Figure~\ref{fig:fig7} illustrates this effect by comparing
 two trajectories, one of a positive muon, the other of a
 negative muon.
 We injected a positive and a negative muon, each of 1~GeV,
 at 20 Km with zenith angle of $11^\circ$ and azimuth of
 $45^\circ$.  The muons were followed 
 until decay in the geomagnetic field at Fort Sumner.
 The narrow dark line is a negative muon deflected in {\em x}
 direction and the wide dark line in {\em y} direction.
 The narrow light line is a positive muon deflected in
 {\em x} direction and
 the wide line a positive deflected in {\em y} direction.
 It is possible to see that the positive muon
 decays before reaching at an altitude of 6000 meters.
 In addition, some muons bend away from the Earth, outside
 the opening angle of the (vertical) muon detector.
 Such variations in the muon track will produce a change
 in the muon fluxes.

\begin{figure*}[ht]
{\centering \begin{tabular}{c}
\includegraphics[angle=0, width=10.0cm]{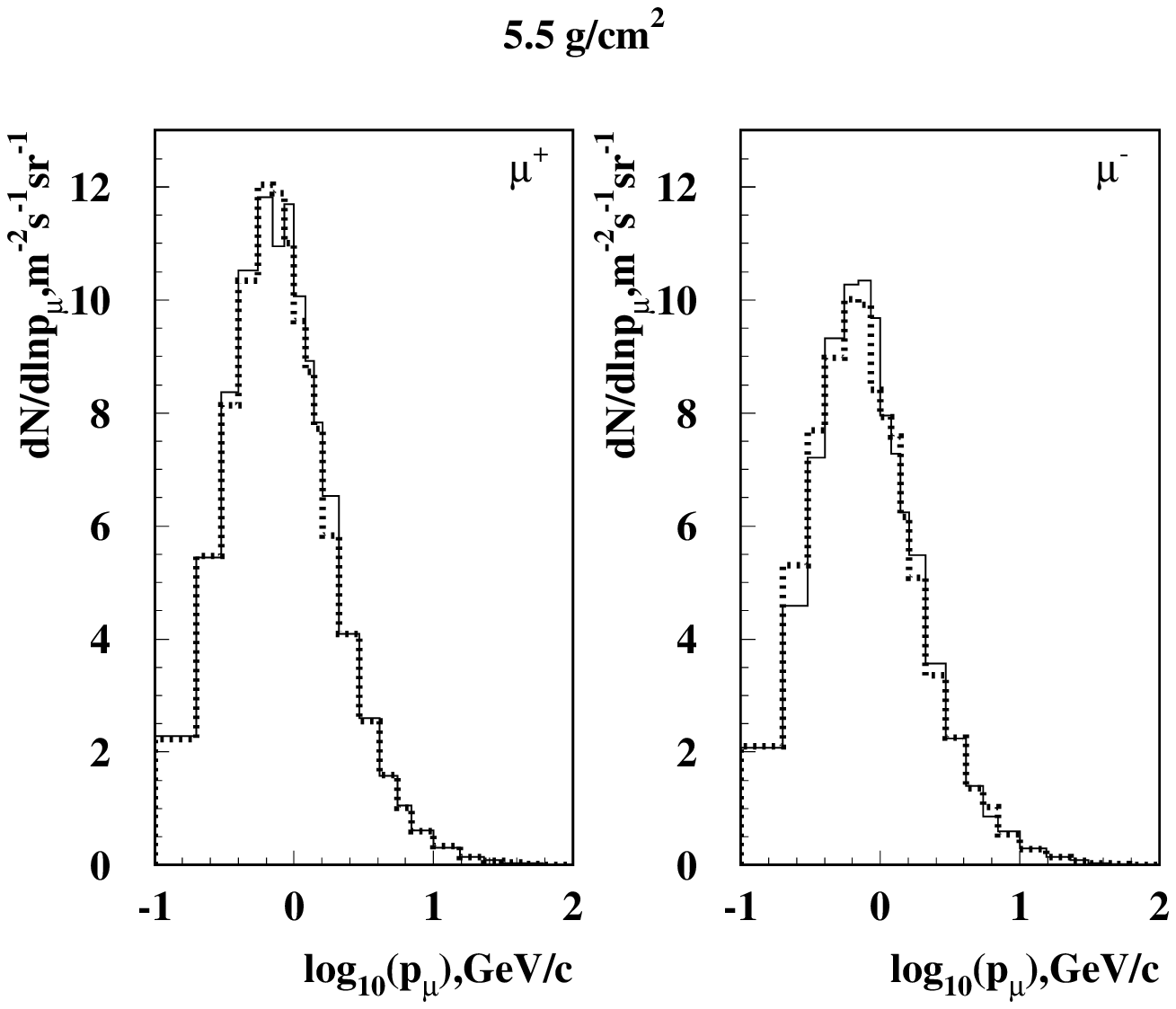}\\
\includegraphics[angle=0, width=10.0cm]{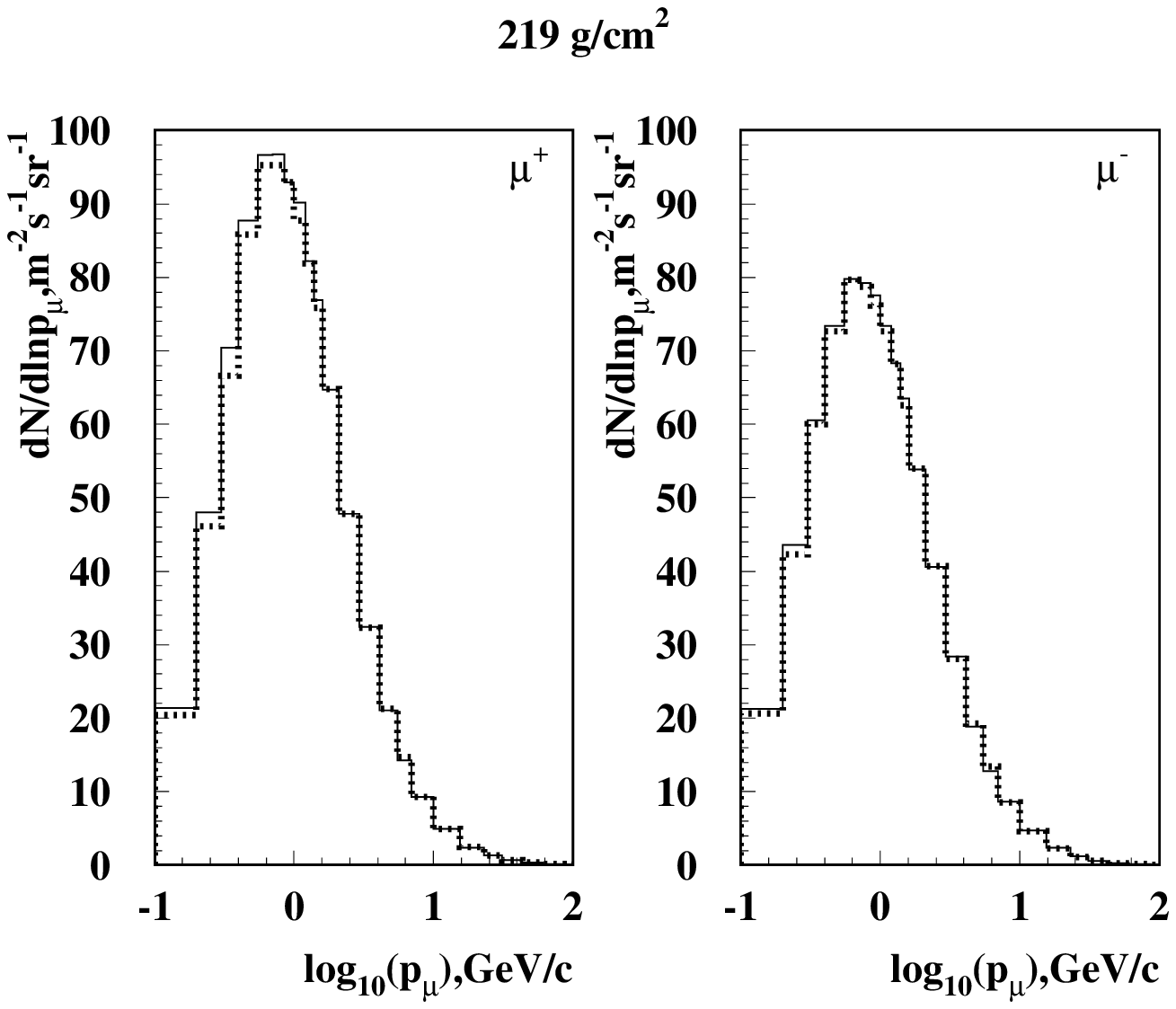}
\\
\end{tabular}\par}
\caption{Solid line: Muon flux with B=0 with the theoretical transmission
function.
Dotted line: Muon flux with B at Fort Summer with the theoretical transmission
function.
}
\label{fig:fig8}
\end{figure*}

\begin{figure*}[ht]
{\centering \begin{tabular}{c}
\includegraphics[angle=0, width=10.0cm]{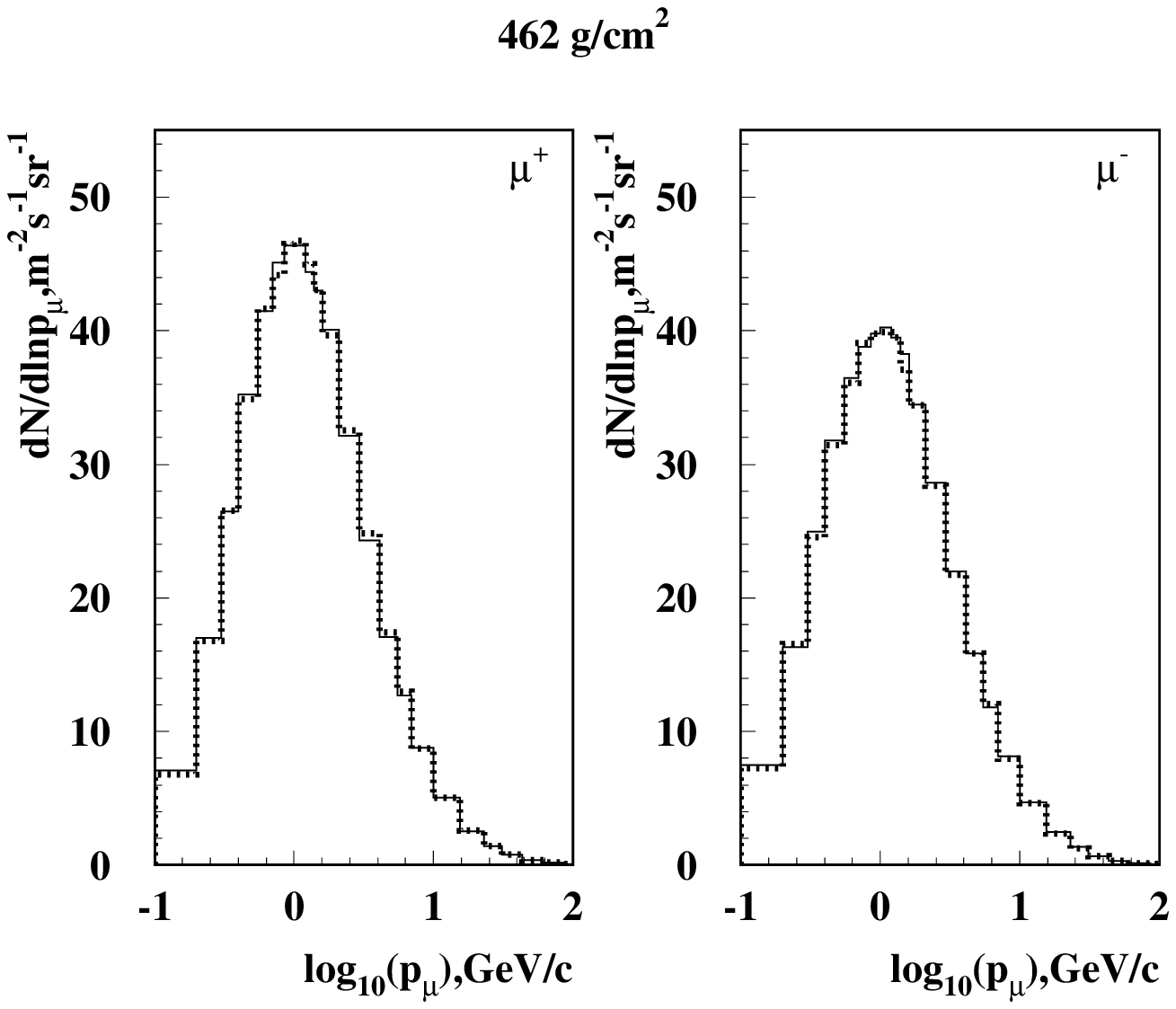}\\
\includegraphics[angle=0, width=10.0cm]{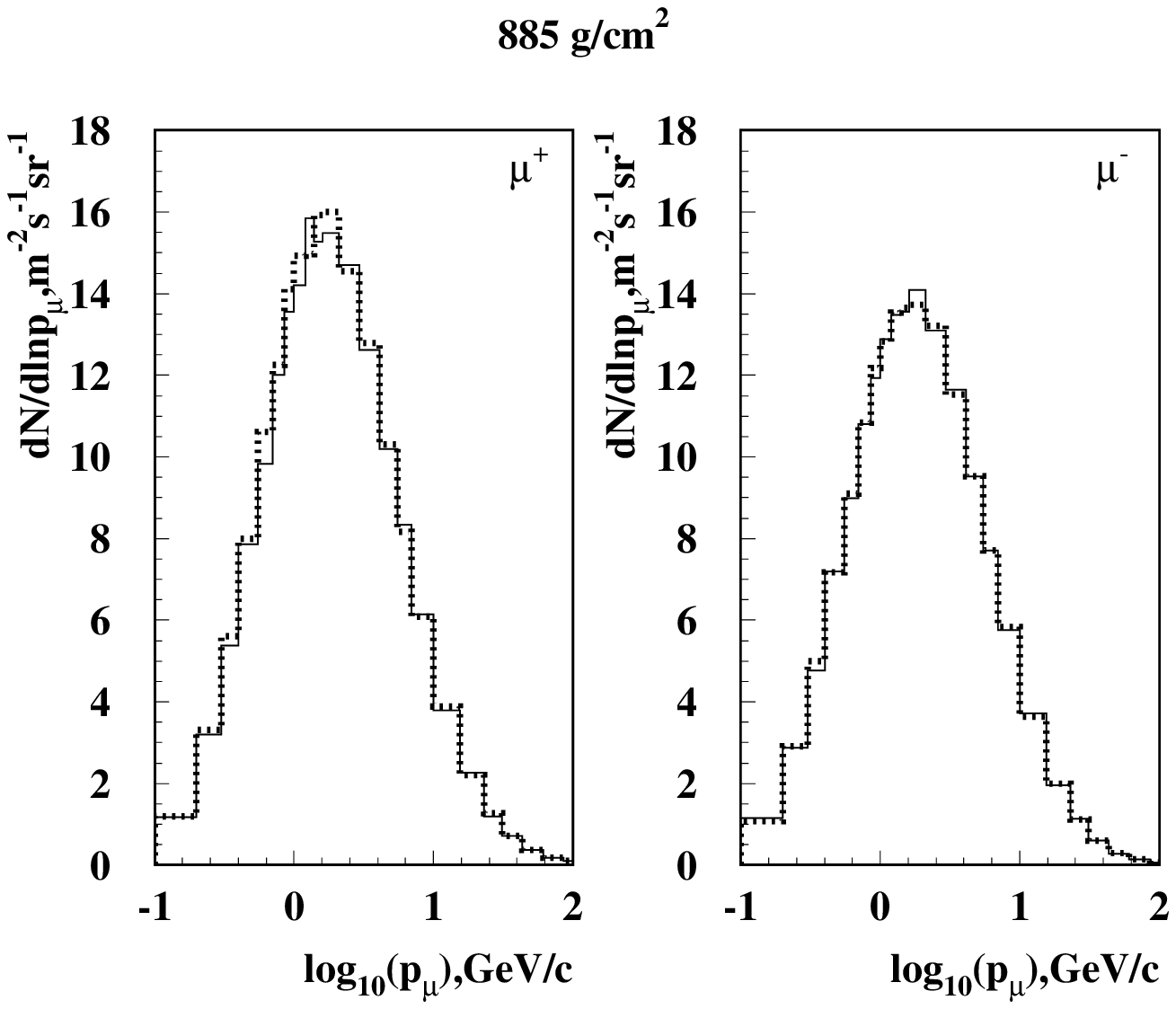}
\\
\end{tabular}\par}
\caption{Solid line: Muon flux with B=0 with the theoretical transmission
function.
Dotted line: Muon flux with B at Fort Summer with the theoretical transmission
function.
}
\label{fig:fig9}
\end{figure*}

 To see this we calculate the flux in function of the momentum 
 with the local magnetic field at Fort Summer and with zero
 magnetic field at different altitudes in the atmosphere.
 In figures~\ref{fig:fig8} and \ref{fig:fig9} it is possible to
 see that the differences are higher at low atmospheric depth
 but there is a remaining difference also at ground level.

\subsection{Muon fluxes from different primary protons
 and helium spectra.
 }

\begin{figure*}[ht]
{\begin{tabular}{c}
\resizebox*{0.5\textwidth}{!}{\includegraphics{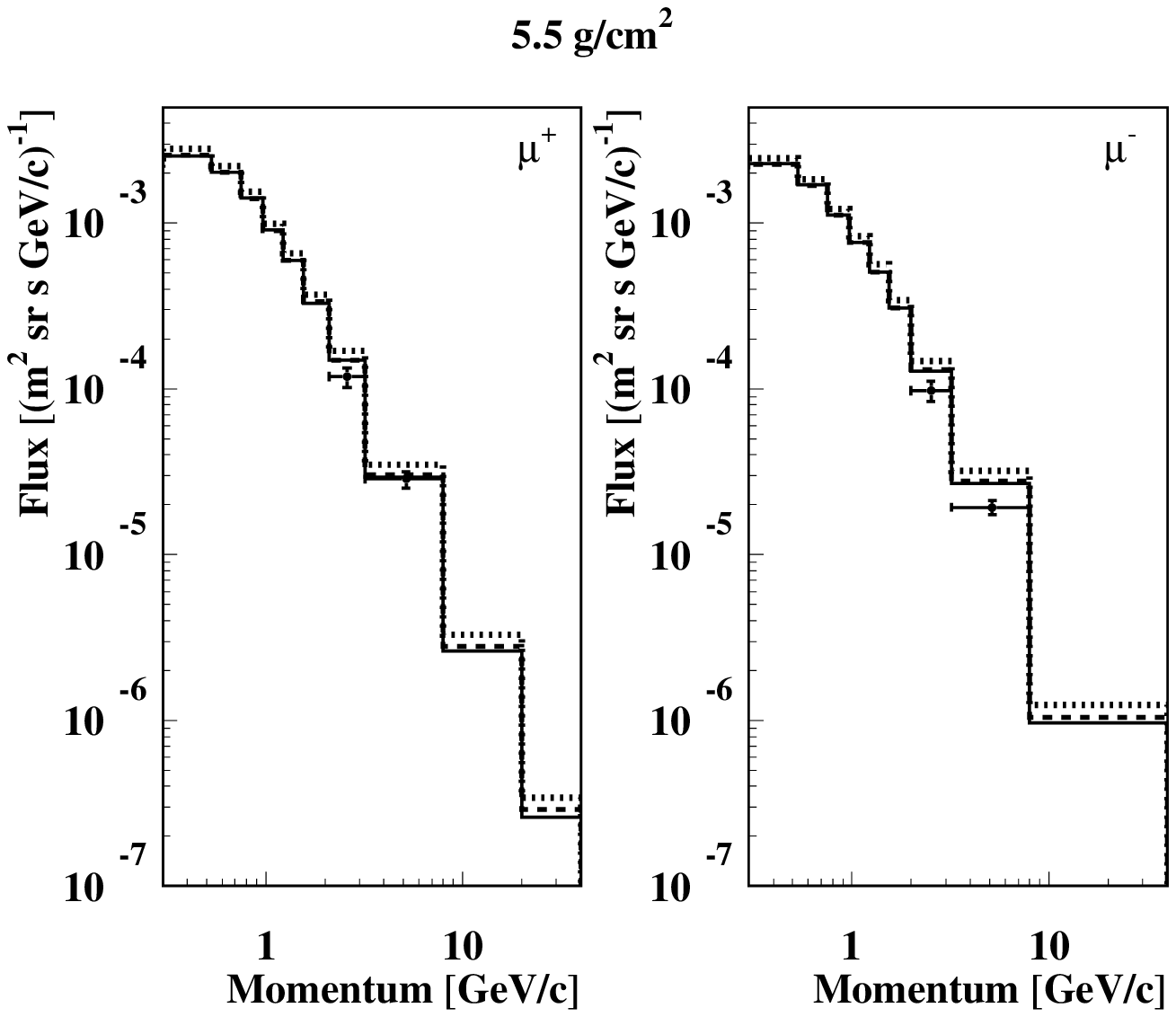}}
\\
\resizebox*{0.5\textwidth}{!}{\includegraphics{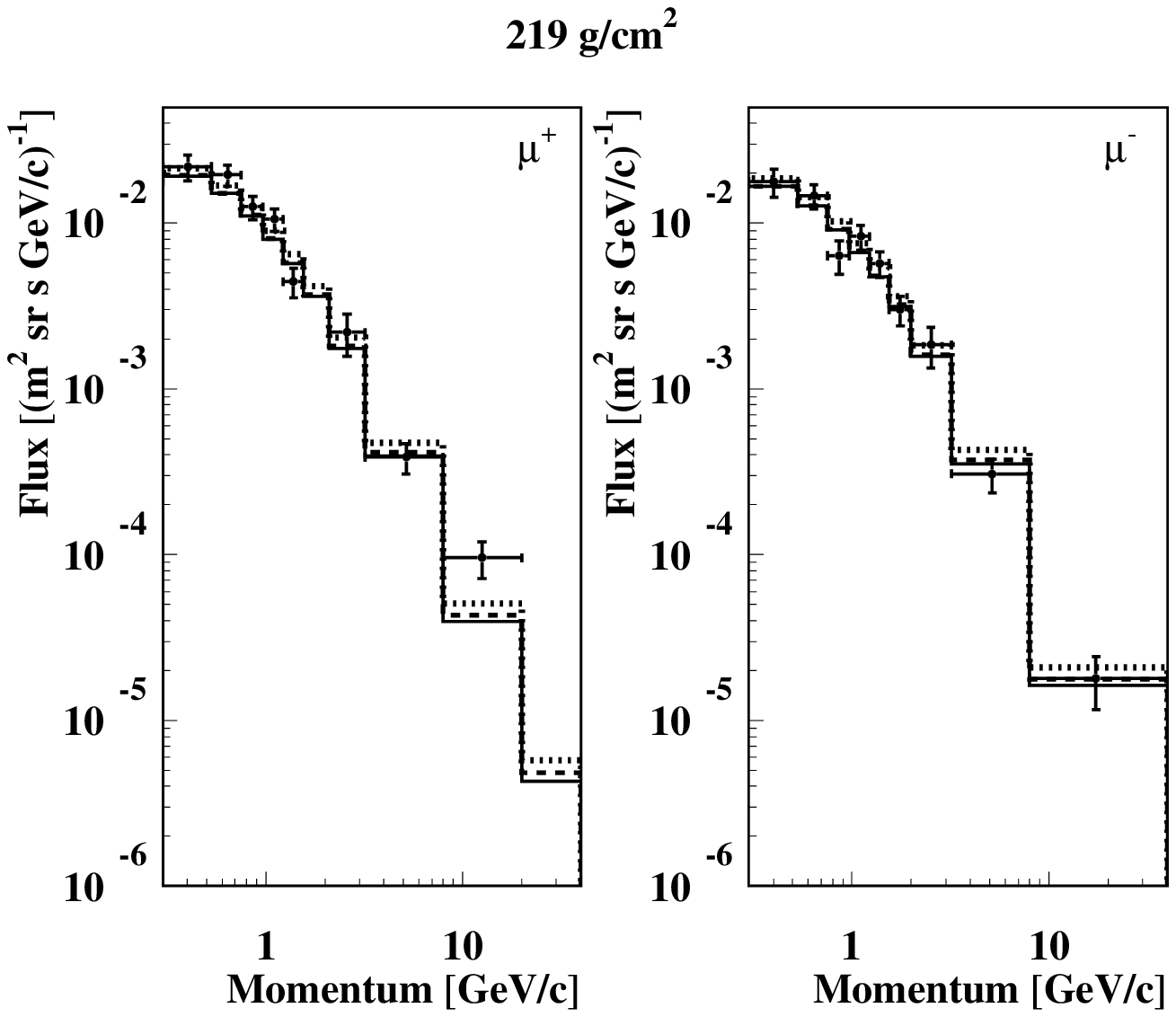}}
\\
\end{tabular}\par}
\caption{
$\mu ^{+}$ and $\mu ^{-}$ fluxes.
Dashed line: using as input Group 1.
Dotted line: using as input Group2.
Solid line: using as input Ref.~\cite{sergioyyo}.
The full circle correspond to the experimental muon
data of CAPRICE98 \cite{yo}.
}
\label{fig:fig10}
\end{figure*}

\begin{figure*}[ht]
{\centering \begin{tabular}{c}
\resizebox*{0.5\textwidth}{!}{\includegraphics{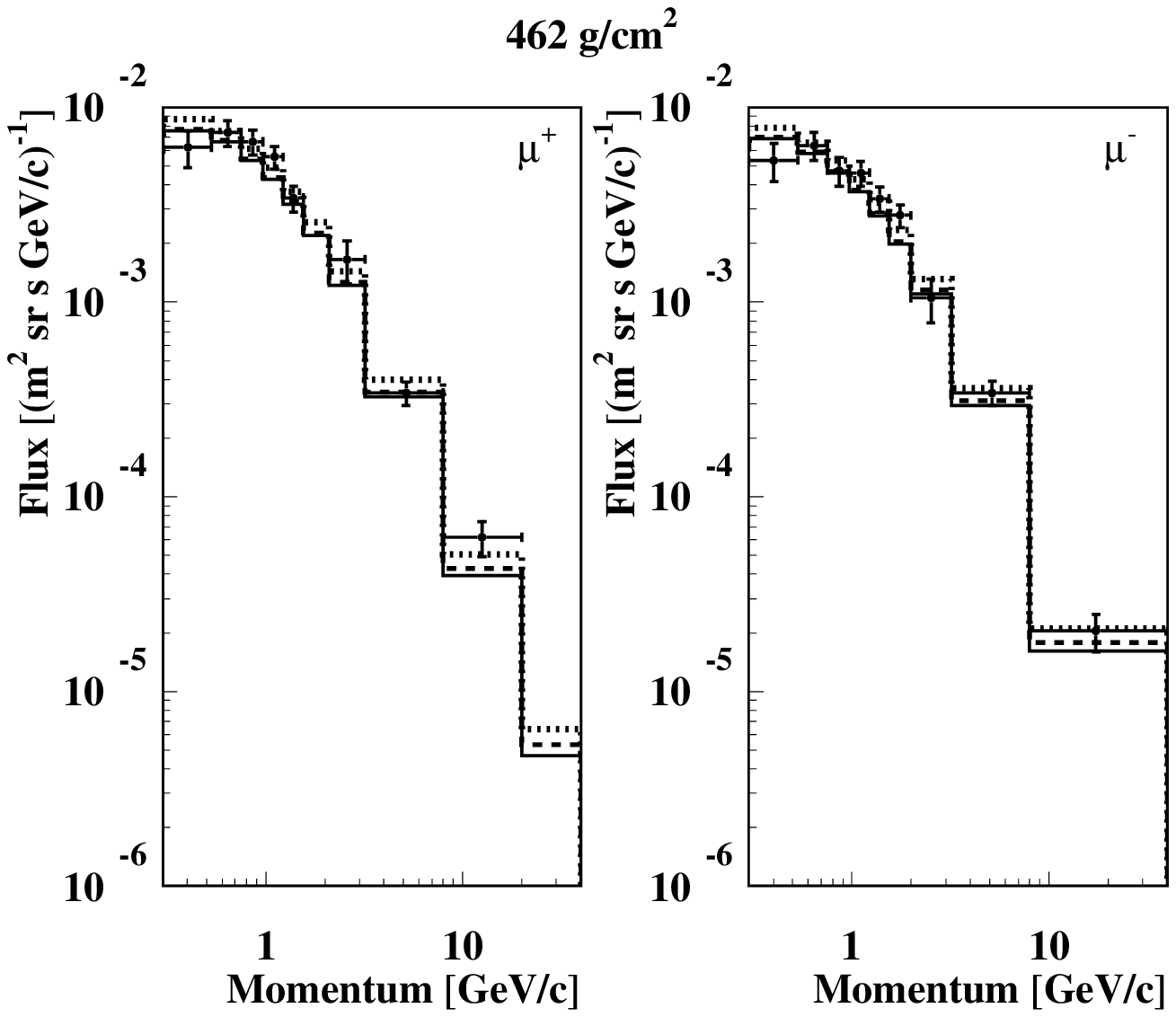}}
\\
\resizebox*{0.5\textwidth}{!}{\includegraphics{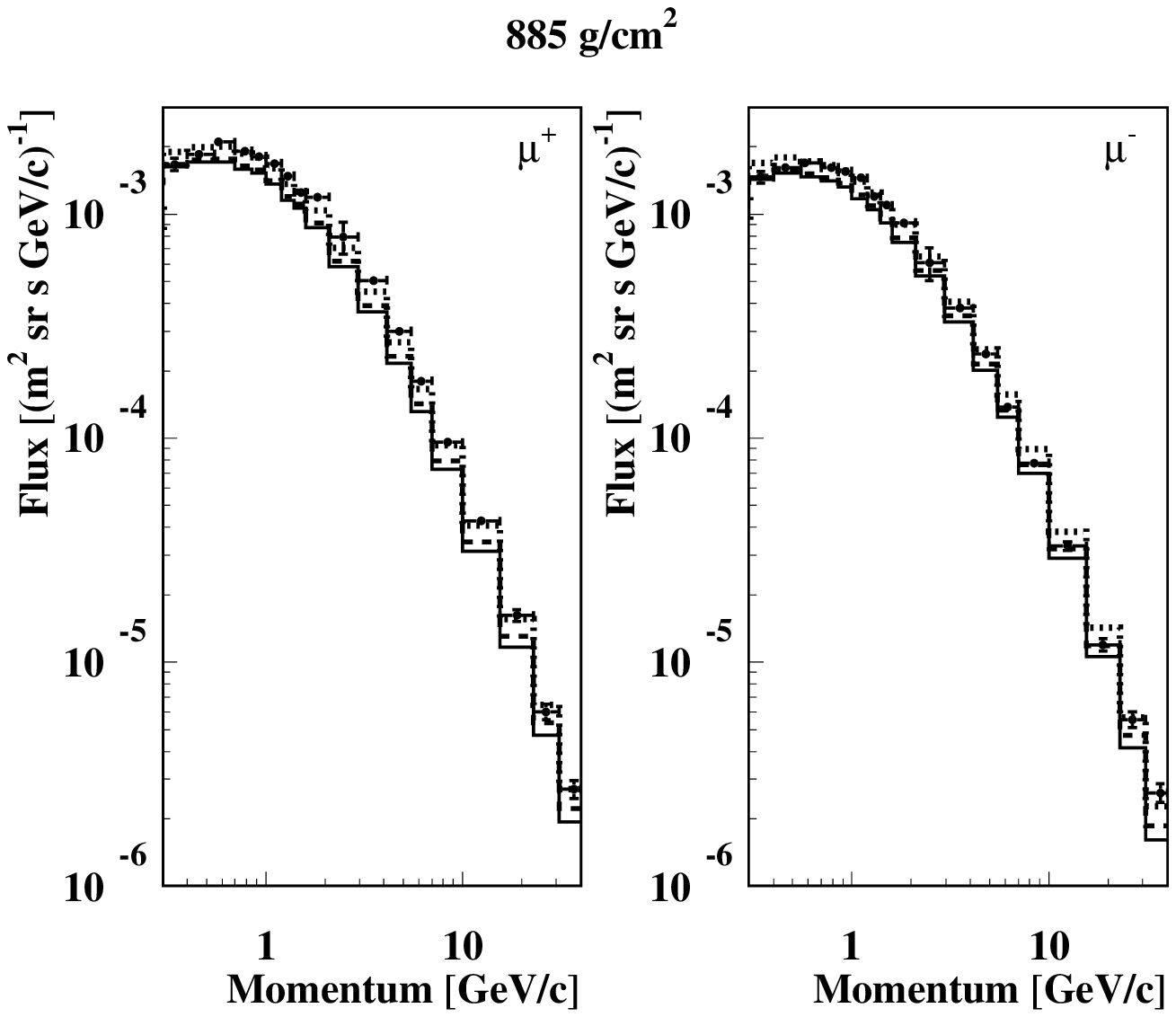}}
\\
\end{tabular}\par}
\caption{
$\mu ^{+}$ and $\mu ^{-}$ fluxes.
Dashed line: using as input Group 1.
Dotted line: using as input Group2.
Solid line: using as input Ref.~\cite{sergioyyo}.
The full circle correspond to the experimental muon
data of CAPRICE98 \cite{yo}.
}
\label{fig:fig11}
\end{figure*}

 We study the uncertainty in the predicted muon fluxes by
 using three different primary spectra of proton and helium
 in the calculation: group 1, group 2 and those from
 Ref.~ \cite{sergioyyo}.
 Figures~\ref{fig:fig10} and \ref{fig:fig11}
 show the fluxes of positive and negative muons
 as a function of momentum using as input Group 1 (dashed line), 
 Group 2 (dotted line) and that of Ref.~\cite{sergioyyo}
 (solid line). The experimental data of the muon flux of CAPRICE98 is
 with solid points.

\begin{figure*}[ht]
{\centering \begin{tabular}{cc}
\resizebox*{0.5\textwidth}{!}{\includegraphics{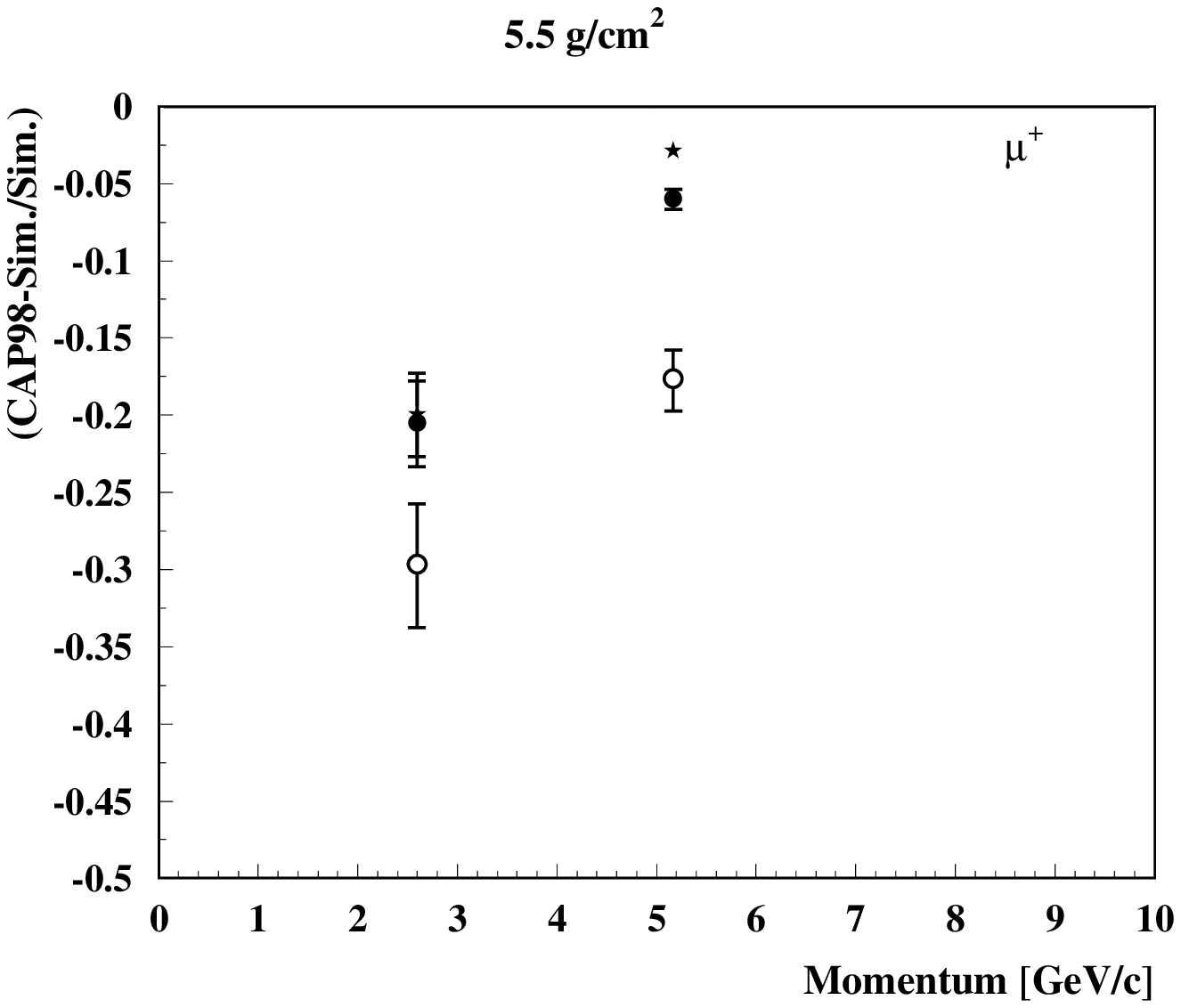}}
&
\resizebox*{0.5\textwidth}{!}{\includegraphics{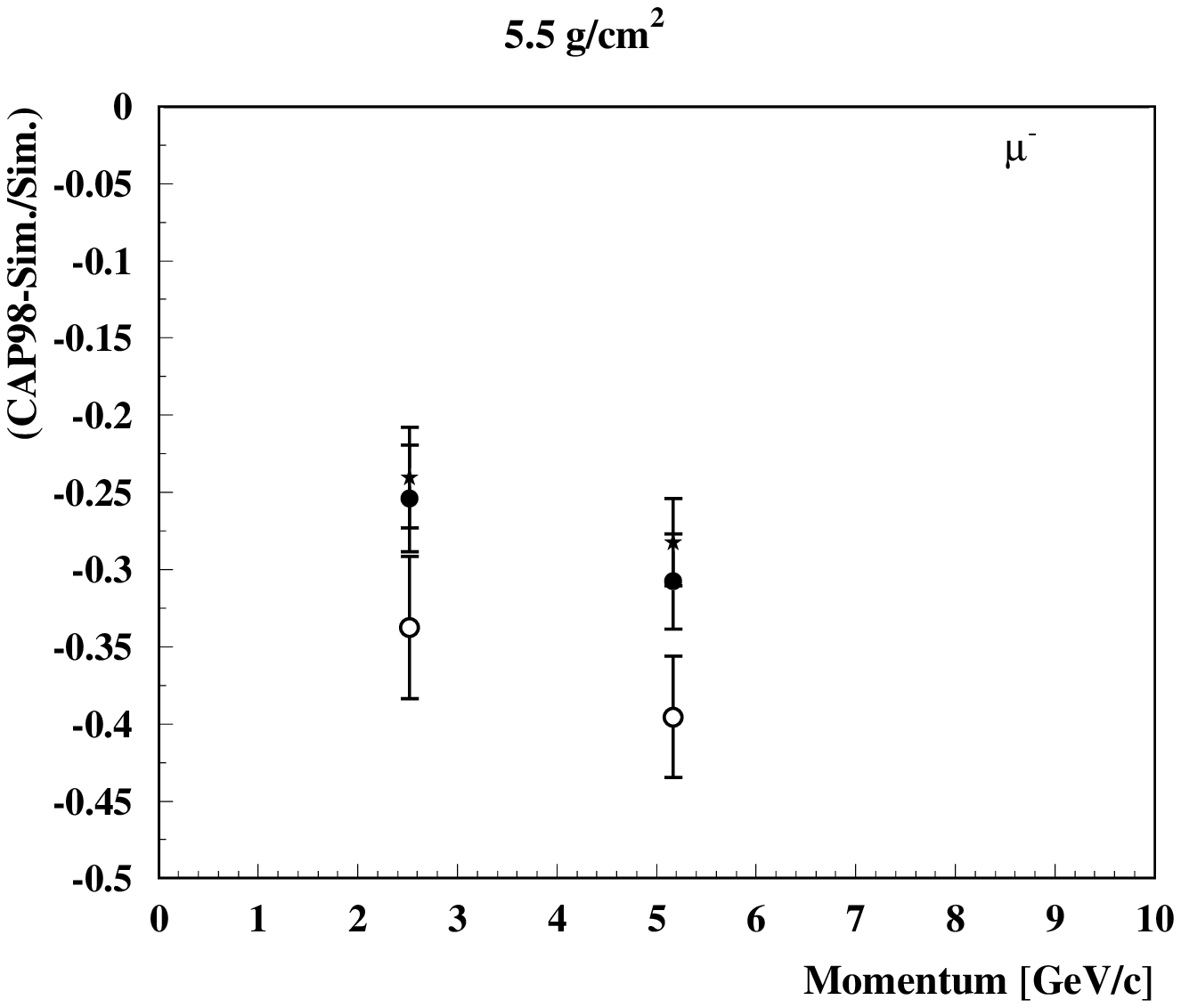}}
\\
\resizebox*{0.5\textwidth}{!}{\includegraphics{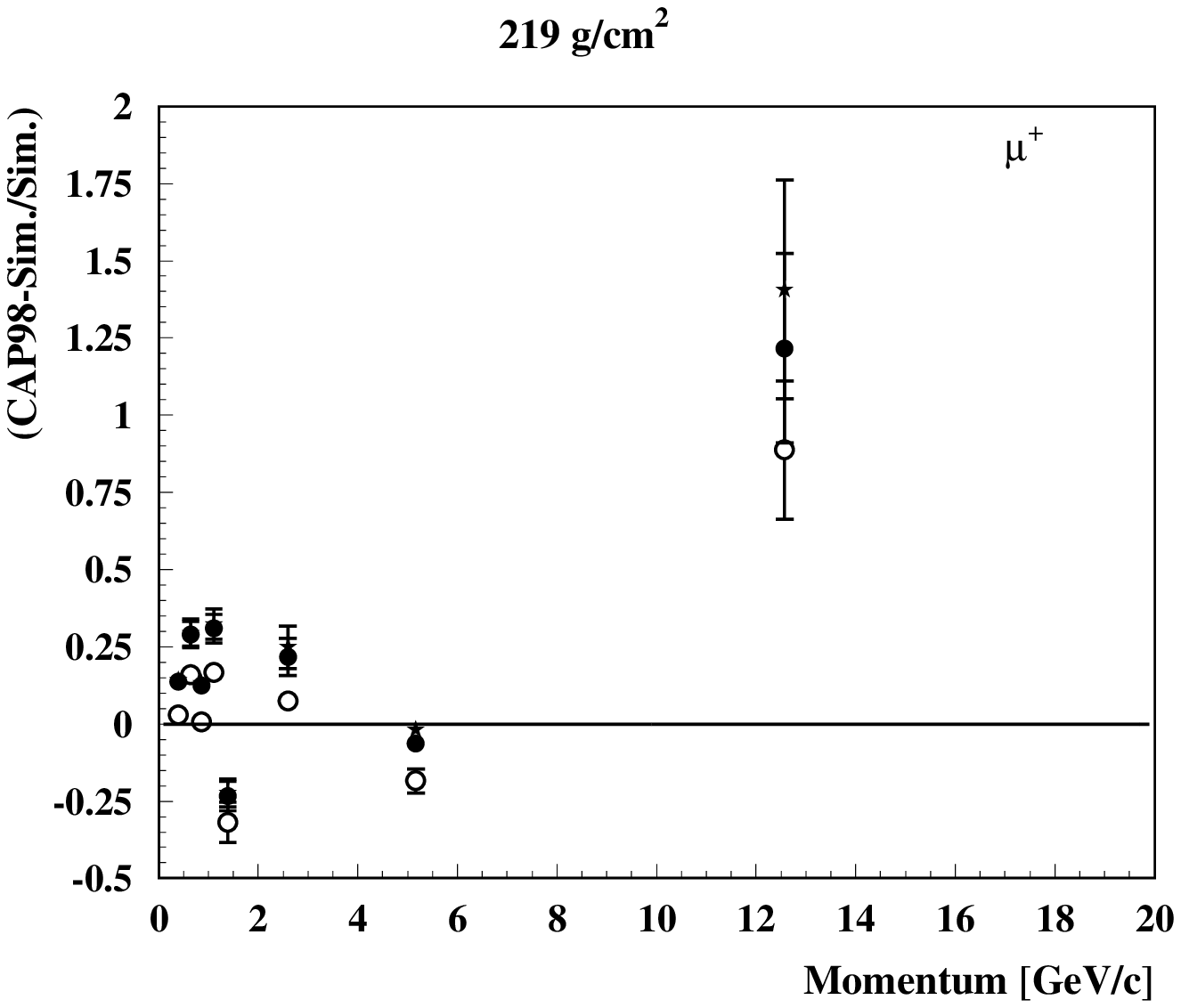}}
&
\resizebox*{0.5\textwidth}{!}{\includegraphics{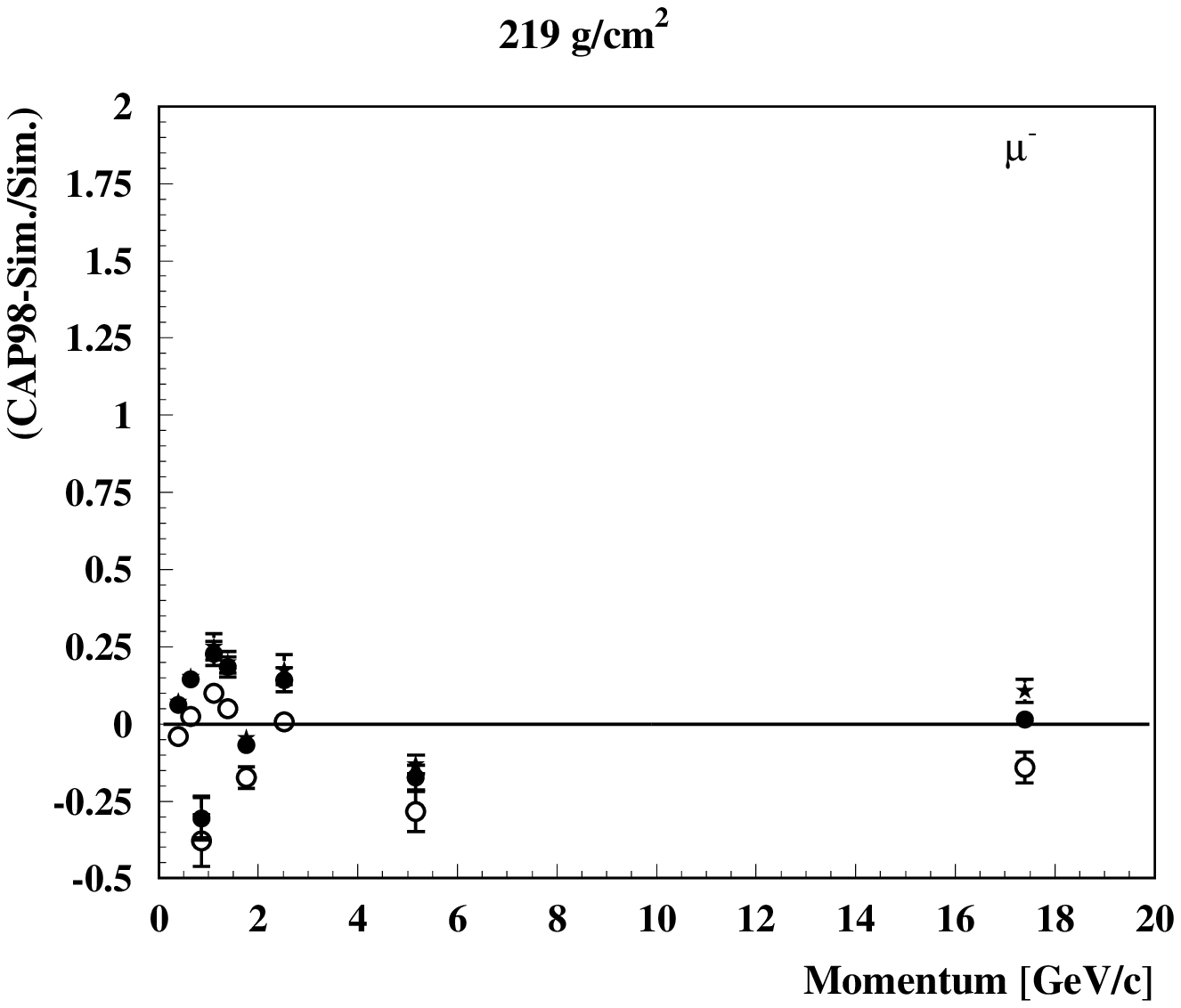}}
\\
\end{tabular}\par}
\caption{
Relative difference between 
$\mu ^{+}$ and $\mu ^{-}$ fluxes obtained from 
caprice 98 experiments and the simulations.
Full circle: using as input Group 1.
Open circle: using as input Group2.
Star: using as input Ref.~\cite{sergioyyo}.
}
\label{fig:fig12}
\end{figure*}

\begin{figure*}[ht]
{\centering \begin{tabular}{cc}
\resizebox*{0.5\textwidth}{!}{\includegraphics{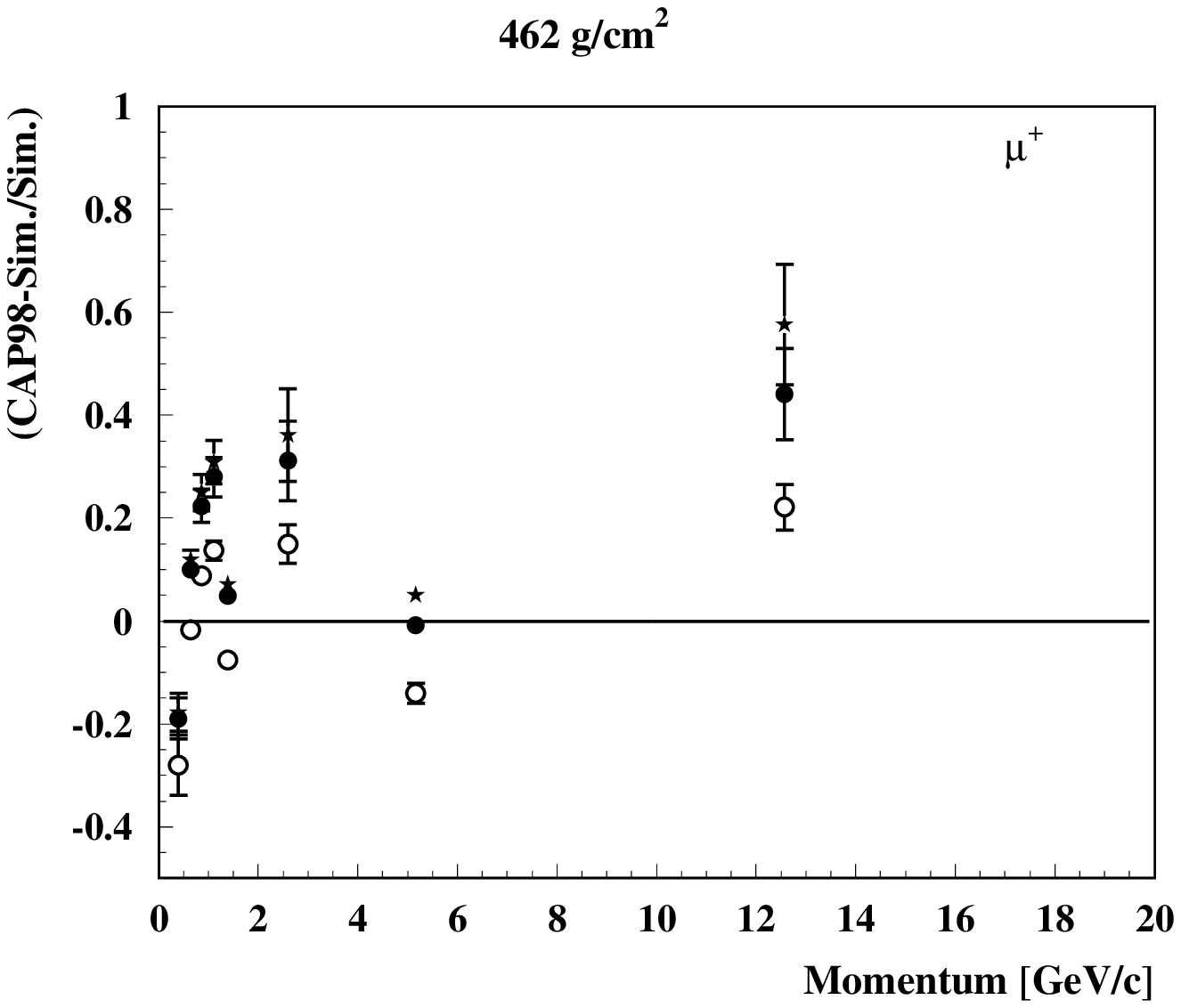}}
&
\resizebox*{0.5\textwidth}{!}{\includegraphics{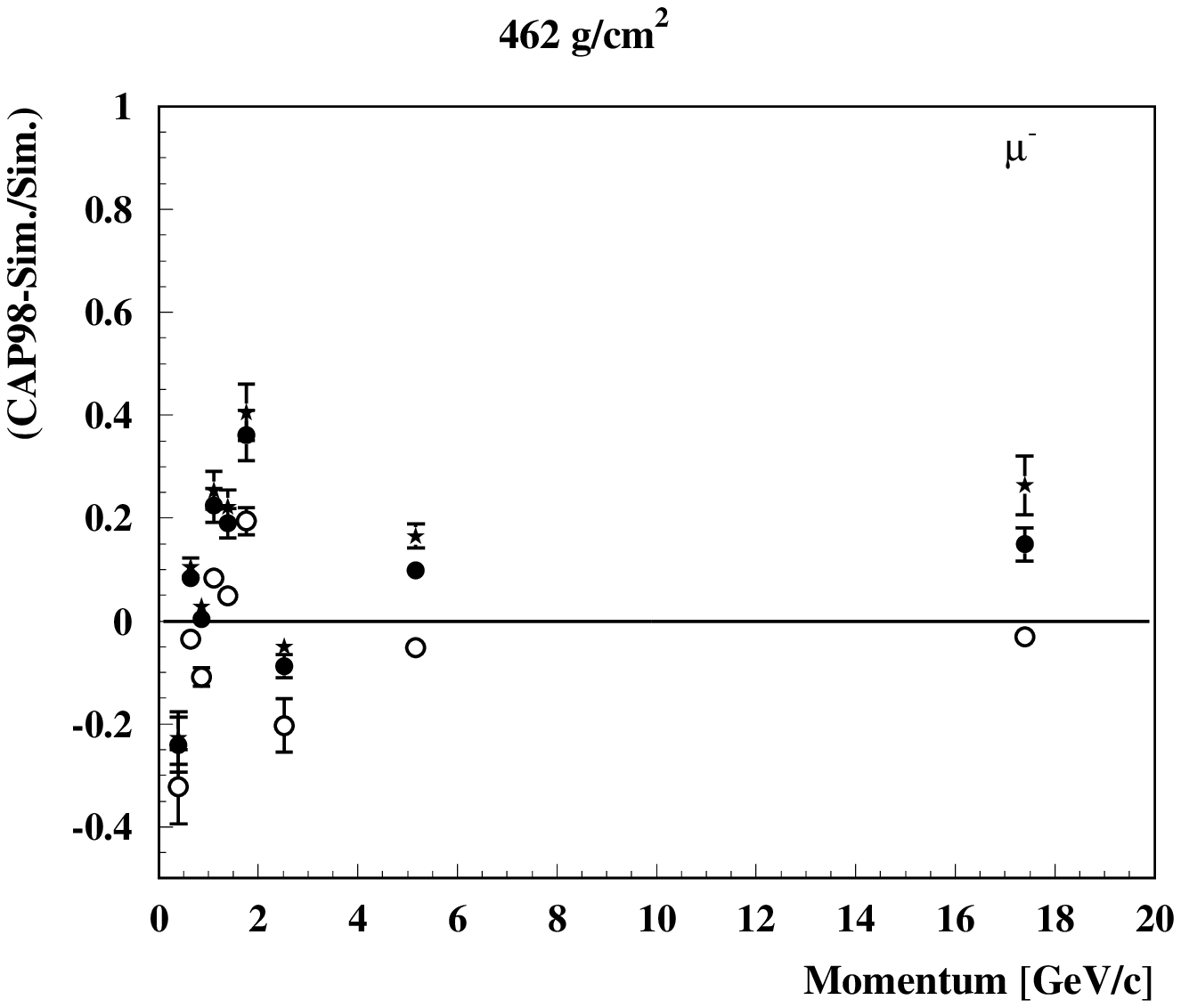}}
\\
\resizebox*{0.5\textwidth}{!}{\includegraphics{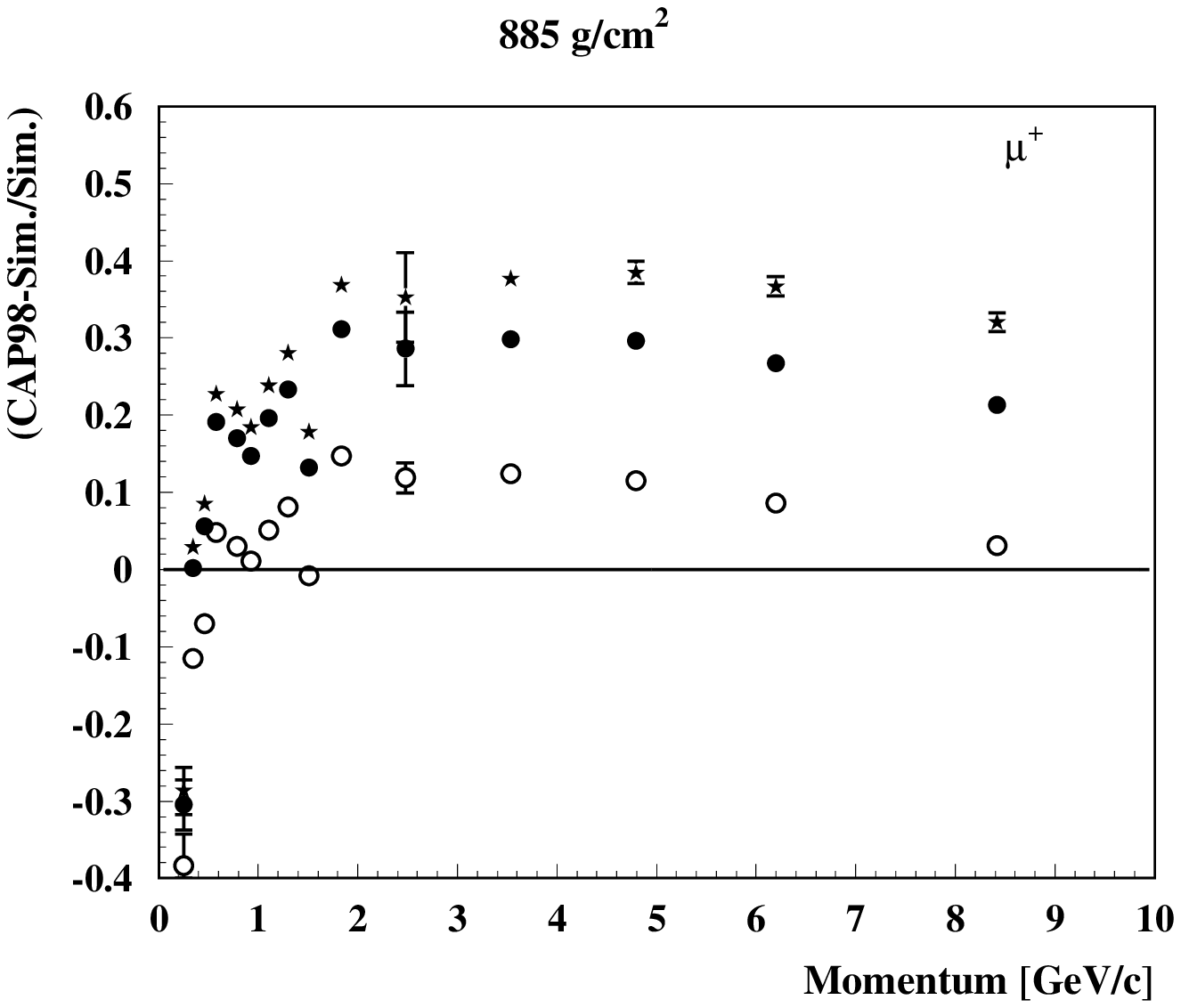}}
&
\resizebox*{0.5\textwidth}{!}{\includegraphics{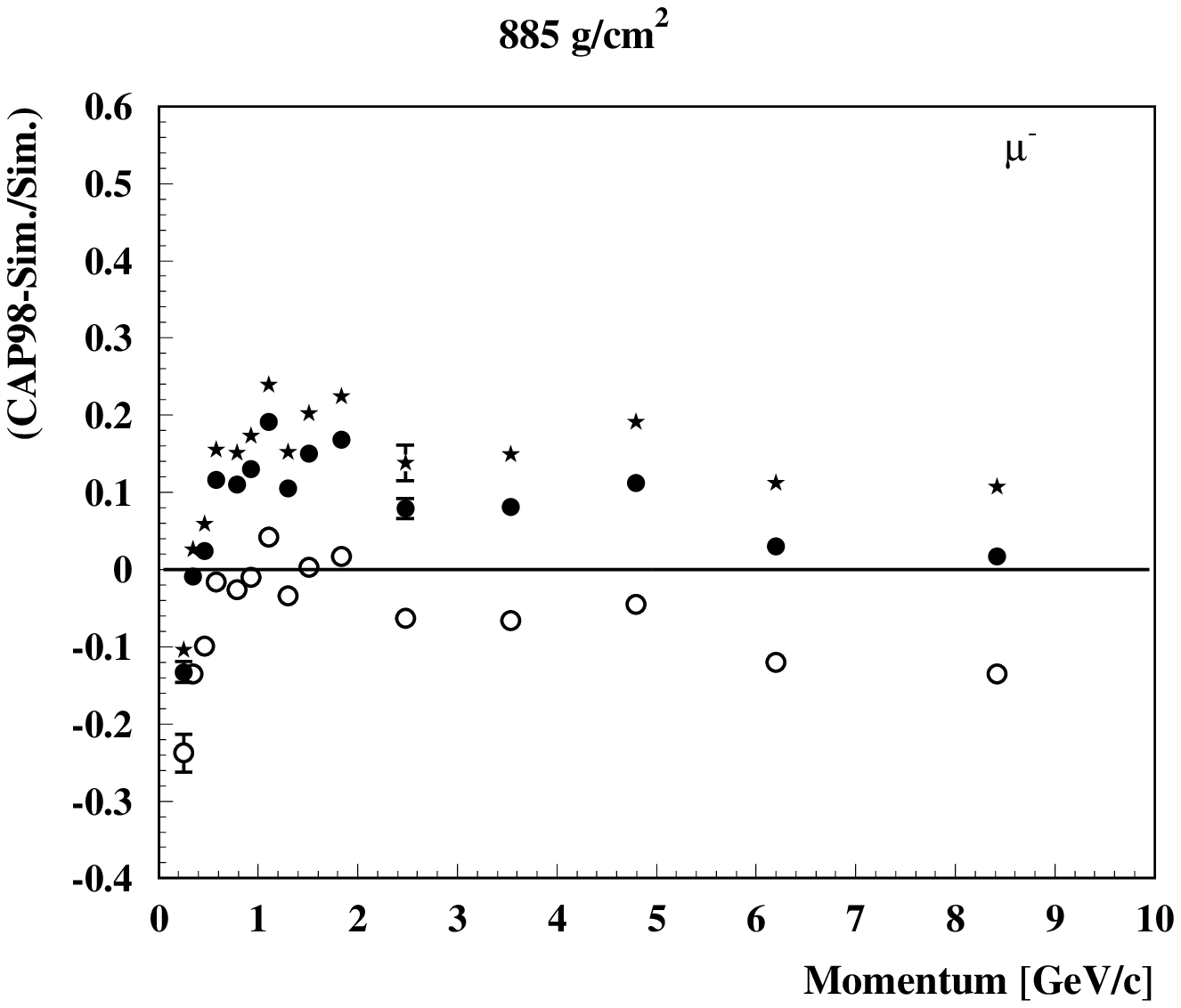}}
\\
\end{tabular}\par}
\caption{
Relative difference between 
$\mu ^{+}$ and $\mu ^{-}$ fluxes obtained from 
CAPRICE98 experiment and the simulations.
Full circle: using as input Group 1.
Open circle: using as input Group2.
Star: using as input Ref.~\cite{sergioyyo}.
}
\label{fig:fig13}
\end{figure*}

 To clarify these plots we calculate the relative difference
 between the CAPRICE98 data and the three results
 of the simulation.
 In figures~[\ref{fig:fig12}-\ref{fig:fig13}] it is possible to see  
 that the higher cosmic ray input fits
 the experimental data somewhat better.
 To quantify the results we also calculate  
 the relative differences between $\mu ^{+}$ and $\mu^-$ fluxes obtained
 from group 1 and group 2 ($\frac{Group2-Group1}{Group1}$).
 At all altitudes the relative difference between the muon fluxes
 is less than 30 \% at energies around 180 GeV and less than 20 \% at
 energies lower  than 20 GeV. The difference in the calculated
 muon charge ratio $\mu ^{+}$ /$\mu ^{-}$ with the different
 inputs is very small, less than 1 \%.

\section{Conclusion}
\label{S:conclusion}

 We have made an analysis of the effect of the geomagnetic field
 outside and inside the atmosphere.
 The differences in the muon fluxes due to the effect of the
 magnetic field outside the atmosphere decrease with increasing
 atmospheric depth. The reason is that muons on the ground are
 generated by higher energy cosmic rays that suffer much less
 from the geomagnetic cutoff.
 This is not the case for the effects due to the local magnetic 
 field. This effect is nearly the same at all altitudes.
 The reason is that muons with higher energy
 do not easily decay and have much longer pathlengths
 that compensate for
 the smaller amount of bending per unit pathlength.

 We also include a calculation of the East-West effect
 and see the difference in the muon fluxes coming from
 these directions. These differences increase at low
 geomagnetic latitude. They are mostly important at float
 altitude.

 We also study the effect of different primary cosmic ray 
 spectra on the predicted muon flux. The best agreement 
 with CAPRICE data that we find is for a higher parametrization
 of the primary cosmic ray flux obtained from the data of BESS
 and JACEE. Differences are most clear at ground level and are
 of order 20\% at low energy and of order 30\% at energies above 100 GeV.
 With the current uncertainties of the primary cosmic ray flux
 it is difficult to make much better prediction of the
 muon fluxes and to normalize the neutrino flux calculation
 to them. We need primary cosmic ray flux data that are
 substantially more accurate than the ones available at present.

\section{Acknowledgments}
 This work is supported in part
 by U.S. Department of Energy contract
 DE-FG02 91ER 40626.

\clearpage

\clearpage
                                                                                
\newpage

\newpage


%
%
%
\end{document}